\documentclass[aps,prb,groupedaddress,twocolumn,10pt,showpacs]{revtex4}
\usepackage[pdftex]{graphicx}
\usepackage{amsmath,amssymb,color,subfigure,yllmath}
\providecommand{\myheading}[1]{\textbf{#1}}

\begin{document}
\title{Bias-free simulation of diffusion-limited aggregation on a square lattice}
\author{Yen Lee Loh}
\affiliation{Department of Physics and Astrophysics, University of North Dakota, Grand Forks, ND  58201, USA}
\date{This version started 2014-6-27; touched 2014-6-27; compiled \today}
\begin{abstract}
We identify sources of systematic error in traditional simulations of the Witten-Sander model of diffusion-limited aggregation (DLA) on a square lattice.
We present an algorithm that reduces these biases to below $10^{-12}$.
We grow clusters of $10^8$ particles on $65536\times 65536$ lattices.
We verify that lattice DLA clusters inevitably grow into anisotropic shapes, dictated by the anisotropy of the aggregation process.
We verify that the fractal dimension evolves from the continuum DLA value, $D=1.71$, for small disk-shaped clusters, towards Kesten's bound of $D=3/2$ for highly anisotropic clusters with long protruding arms.
\end{abstract}
%Computer modeling and simulation, 07.05.Tp
%Computational techniques
% statistical physics and nonlinear dynamics, 05.10.-a
% disordered solids, 61.43.Bn
%Fractals
%structure of disordered solids, 61.43.Hv
%Structure
%of disordered solids, 61.43.-j
%of fractals, 61.43.-j
%74.62.En	Effects of disorder
\maketitle

Diffusion-limited aggregation (DLA) is one of the most important models in nonequilibrium statistical physics, exhibiting self-organized criticality\cite{bak1987} and complex pattern formation.
In the DLA process, one begins with a cluster (``seed crystal'') immersed in a very dilute solution of particles (``molecules'').  Each particle wanders around according to Brownian motion until it encounters the cluster, at which point it ``freezes'' and becomes part of the cluster.  It is more likely that a diffusing particle will stick to a protrusion on the cluster than to a depression.  Thus DLA has a natural instability resulting in pattern formation; protrusions grow quickly and spawn other protrusions, forming a treelike pattern somewhat like frost on a window pane.
DLA and its variants have been used to model a whole host of nonequilibrium phenomena, including 
viscous fingering 	(pattern formation when one fluid is injected into a viscous fluid), 
electrodeposition,\cite{brady1984}
dielectric breakdown,
and surface poisoning in ion-beam microscopy.\cite{halsey2000}

During the course of a DLA simulation, the radius of gyration of the cluster, $R$, and the cluster mass, $M$, are usually recorded.  These data can usually be fit to a power law $R \sim M^\beta$, where $\beta$ is known as the radius-of-gyration exponent.  Then one has $M \sim R^D$ where $D=1/\beta$ is the fractal dimension of the DLA cluster $D$ (insofar as the fractal dimension can be defined for an inhomogeneous finite object).  
%A two-dimensional (2D) DLA cluster has a fractal dimension between $D=1$ (as for a straight line) and $D=2$ (as for a solid disk).  
In two dimensions (2D), $\beta \leq 2/3$; this is one of the few rigorous results\cite{kesten1987} on DLA.
Other results have been obtained using mean-field theories\cite{brener1991,levine1992} and iterated conformal maps,\cite{hastings1997,hastings1998} but these are uncontrolled approximations without small parameters.\cite{halsey2000}  
The bulk of the literature involves numerical simulations.
Early papers\cite{witten1981,witten1983,meakin1983pra2,meakin1983pra3} reported similar values ($\beta\approx 0.585$) for 2D continuum, square lattice, and triangular lattice DLA clusters.
Later papers\cite{meakin1983prl,ball1985,ball1985jpa,meakin1987,halsey2000,menshutin2011} claimed that square lattice DLA clusters evolve from a roughly circular shape for small clusters towards diamond and cross shapes for larger clusters, and that $\beta$ approaches $2/3$ for very large clusters.
Some numerical data suggested that DLA exhibits multiscaling,\cite{coniglio1989,coniglio1990,menshutin2006}  but later work suggested that multiscaling is a finite-size effect that is not intrinsic to DLA.\cite{menshutin2012}

In Monte Carlo simulations of equilibrium systems, such as Ising models, it is essential to construct a Markov chain with the correct invariant distribution (e.g., by ensuring detailed balance).  Any bias in the simulation results in sampling the wrong probability distribution, giving wrong answers for thermodynamic quantities and critical exponents. 
Non-equilibrium situations such as DLA deserve the same amount of care.
However, DLA studies to date have used approximations with errors potentially as large as $1\percent$.  The results were justified by noting that varying the severity of the approximation did not noticeably affect the results; nevertheless one may still be concerned that the approximations led to subtle effects that went unnoticed.

In this paper we present an algorithm for square lattice DLA where probability distributions are sampled with accuracies better than $10^{-12}$.  We verify that DLA on a lattice produces anisotropic clusters, and that the anisotropy originates from the aggregation process rather than the diffusion process.  We confirm that small circular clusters have a radius-of-gyration exponent $\beta=0.585=1/1.71$, but as they mature into anisotropic shapes with arms extending outwards, the exponent tends towards $\beta=0.667=2/3$.

%============================================================================
\section{Biases in traditional DLA}
%============================================================================
The standard algorithm for square lattice DLA\cite{witten1981,witten1983} is as follows:

\begin{enumerate}
\item 
Let the initial cluster consist of a single seed particle at the origin of the lattice.
\item 
Launch a new particle on a launching circle of radius $R_L$ that contains the current cluster.  In other words, generate an angle $\phi$ from the uniform distribution on $[0,2\pi)$, and set $x=\text{round} (R_L \cos \phi)$ and $y=\text{round} (R_L \sin \phi)$.
\item 
Move the particle east, west, north, or south with equal probability.
\item 
If the particle is adjacent to the cluster, add it to the cluster, and go back to step 2. 
\item 
If the particle has diffused outside the killing circle of radius $R_K$, discard it, and go back to step 2.  
\item 
Go back to step 3.
\end{enumerate}

The launching radius is typically taken to be $R_L=R_B+5$ where $R_B$ is the radius of the bounding circle circumscribing the cluster.\cite{witten1981,witten1983,meakin1983pra3}
The killing radius may be as small\cite{witten1981,witten1983} as $R_K=2R_B$ or as large\cite{ball1985jpa} as $R_K=100R_B$.

Obviously, DLA is a stochastic process involving random numbers.  Measurements of  observables (such as $D$) are subject to random error, which cannot be eliminated, but can be reduced by averaging over many simulations, or by self-averaging as part of going to larger system sizes.  
However, one should be wary of systematic errors.  These cannot be removed by any amount of statistical averaging.  Moreover, emergent phenomena such as the self-organized critical behavior of DLA may be strongly affected by any bias inadvertently introduced by the algorithm.  The original algorithm suffers from two potential sources of systematic error:

\begin{enumerate}
\item 
The launching circle only passes through a few lattice points.  When launching a new particle, we must snap its coordinates to the grid, introducing roundoff error.  For a particle accreting onto a cluster of linear size $10^2$ one may worry that the errors may be as large as $10^{-2}$.
\item
For a 2D random walk, even if a particle has wandered outside the killing circle, there is a $100\percent$ probability that it will eventually re-enter the launching circle.  
The particle is more likely to enter at the near side of the circle than at the far side.  By removing the particle from the killing circle and re-launching it from a uniform distribution on the launching circle, the algorithm introduces a bias that may affect results such as $D$.  Even if the killing radius is $10^2$ times the cluster radius, the errors in the return probabilities may still be as large as $10^{-2}$.
\end{enumerate}

%============================================================================
\section{Eliminating launching bias\label{secLaunchingAnnulus}}	
%============================================================================
%----------------------------------------------------------------------------
\myheading{Snap-to-grid error due to launching circle:}
%----------------------------------------------------------------------------
Let us first address the first source of systematic error.  Suppose we launch a particle on a launching circle of radius $R_L$ and snap its coordinates to the grid as described earlier.  The probability distribution of the point $(x,y)$ is
	\begin{align}
	P_{x_0y_0} &=
		\int_{x_0-\half}^{x_0+\half} dx~
		\int_{y_0-\half}^{y_0+\half} dy~
		\delta\big( \sqrt{x^2+y^2} - R_L \big)  
		.
	\label{eqSharpCircleProbabilities}
	\end{align}
Suppose we generate many particles from this distribution and let them diffuse via Brownian random walks.  What is the steady-state concentration of particles within the launching circle?  How far does it deviate from a uniform distribution?

This Brownian problem maps to an electrostatics problem.  The source distribution maps to a charge distribution $Q_{x_0y_0} = P_{x_0y_0}$, and the steady-state distribution of particles maps to the electric potential
	\begin{align}
	V_{xy} &= \sum_{x'y'} G_{x-x',y-y'} Q_{x'y'}
	\end{align}
where $G_{xy}$ is the Green function of the square lattice Poisson equation such that
	\begin{align}
	4G_{xy} - G_{x+1,y} - G_{x-1,y} - G_{x,y+1} - G_{x,y-1} = \delta_x \delta_y 
	.
	\end{align}
Because of the long-range logarithmic divergence in 2D, $G_{xy}$ contains an infinite additive constant.  Thus we define the regularized Green function
$
	F_{xy} = G_{00} - G_{xy}
$.
The quantities $F_{xy}$ are related to the resistances between two points on a square lattice of resistors,\cite{kleinertBook,atkinson1999,cserti2000} and as described in Appendix \ref{secGreenFunction}, it can be calculated to machine precision for any $x$ and $y$.  Compare the potential with that at a reference point $(x'',y'')$, which might as well be the origin $(0,0)$:
	\begin{align}
	V_{xy} - V_{x''y''} &= - \sum_{x'y'} (F_{x-x',y-y'} - F_{x''-x',y''-y'}) Q_{x'y'}
	.
	\end{align}

\begin{figure*}[!htb]
	\subfigure[]{
		\includegraphics[width=.65\columnwidth]{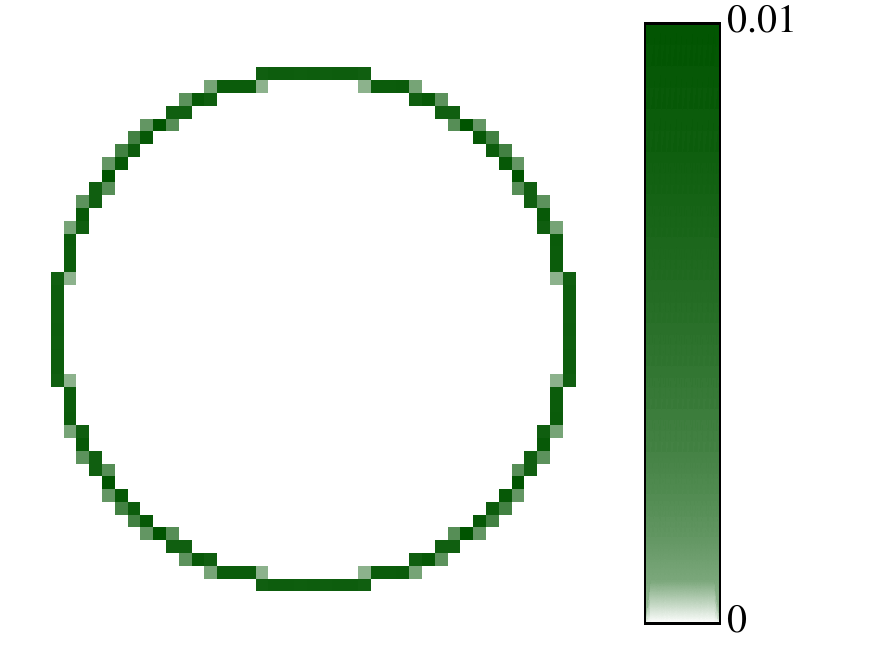}
		\label{CircleCharges}
	}
	\subfigure[]{
		\includegraphics[width=.65\columnwidth]{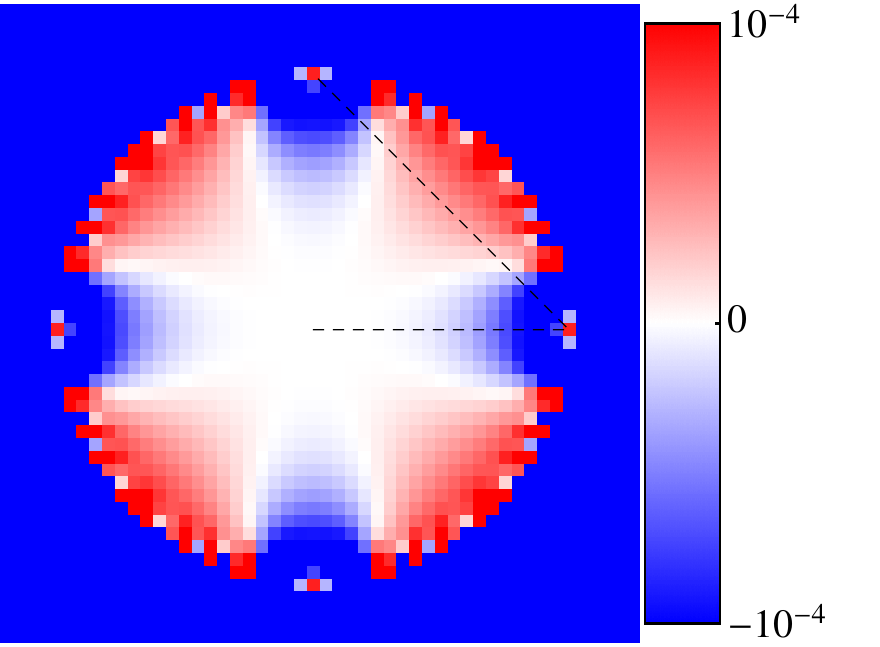}
		\label{CirclePotentials}
	}
	\subfigure[]{
		\includegraphics[width=.6\columnwidth]{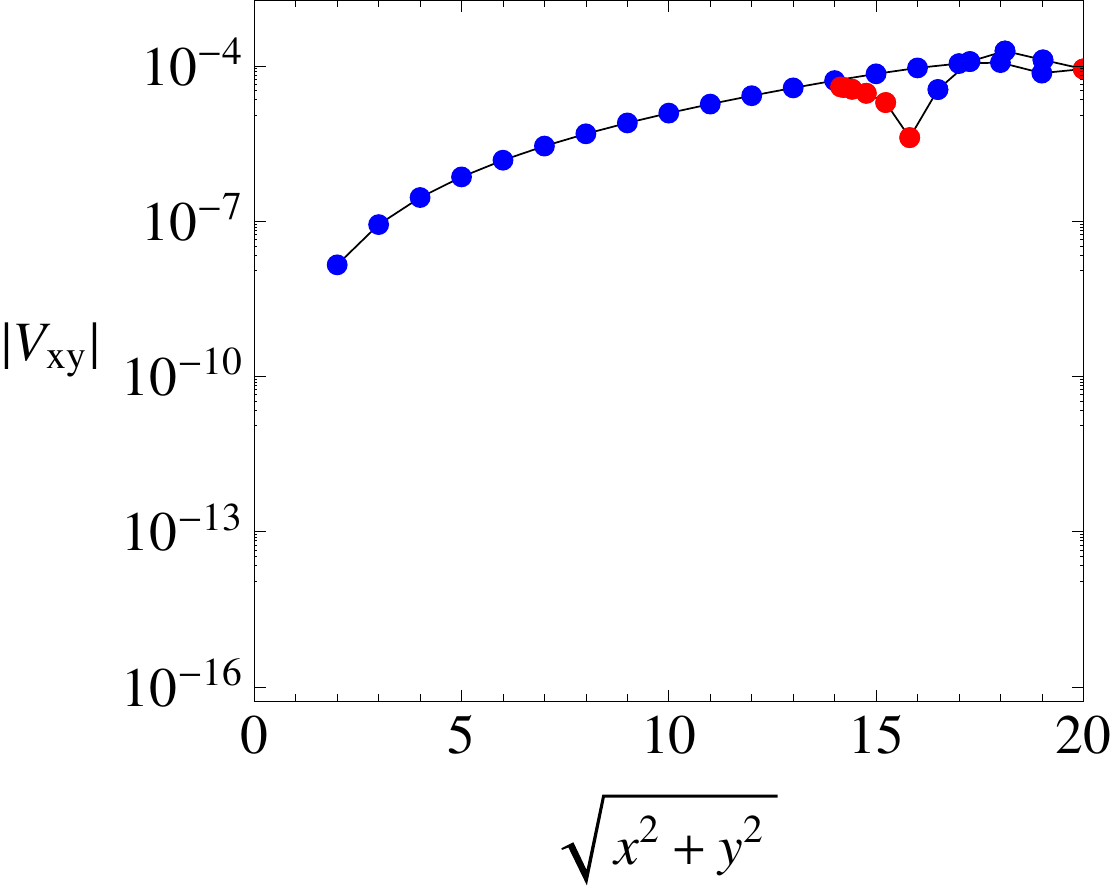}
		\label{CirclePotentialsAlongPath}
	}
	\subfigure[]{
		\includegraphics[width=.65\columnwidth]{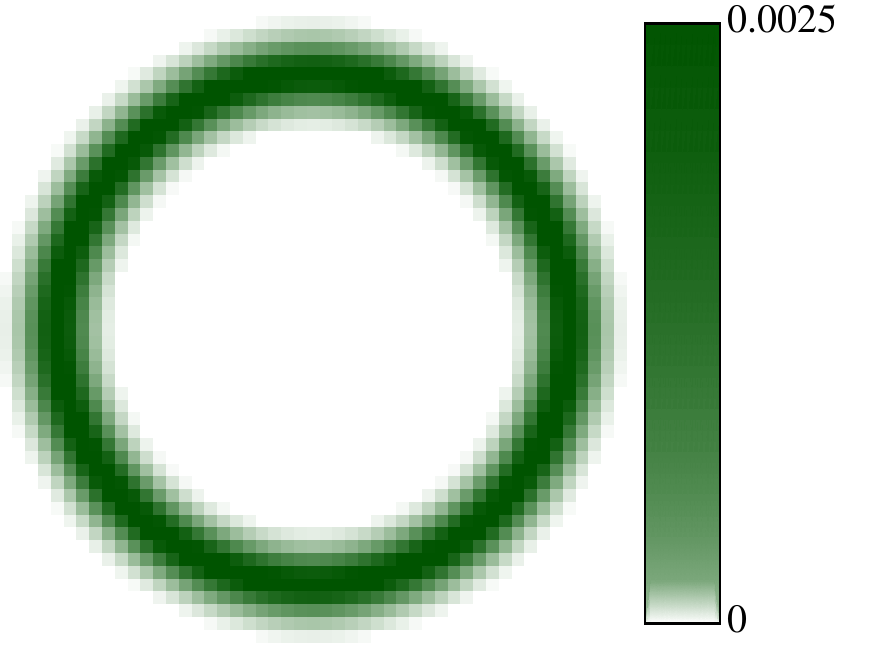}
		\label{AnnulusCharges}
	}
	\subfigure[]{
		\includegraphics[width=.65\columnwidth]{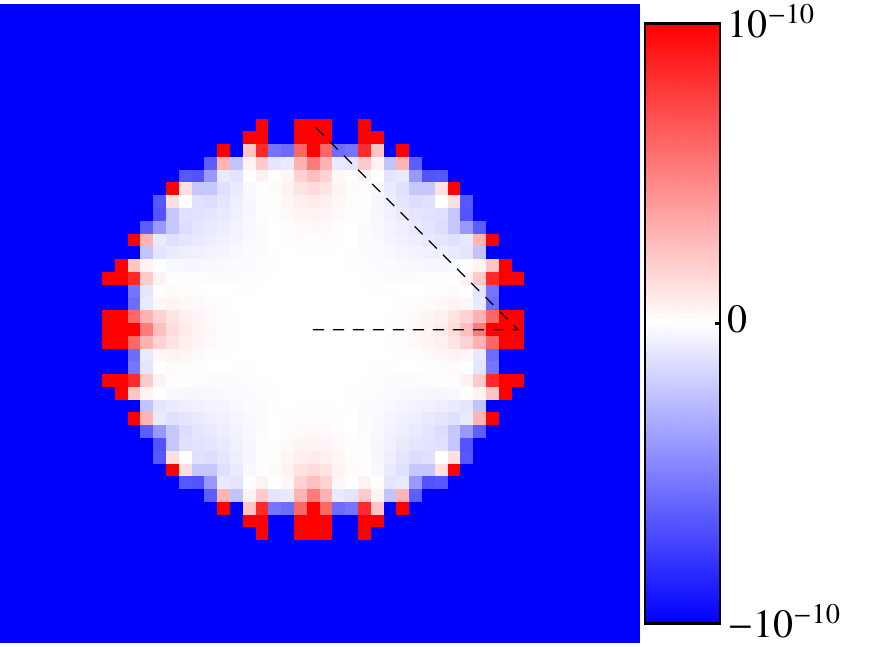}
		\label{AnnulusPotentials}
	}
	\subfigure[]{
		\includegraphics[width=.6\columnwidth]{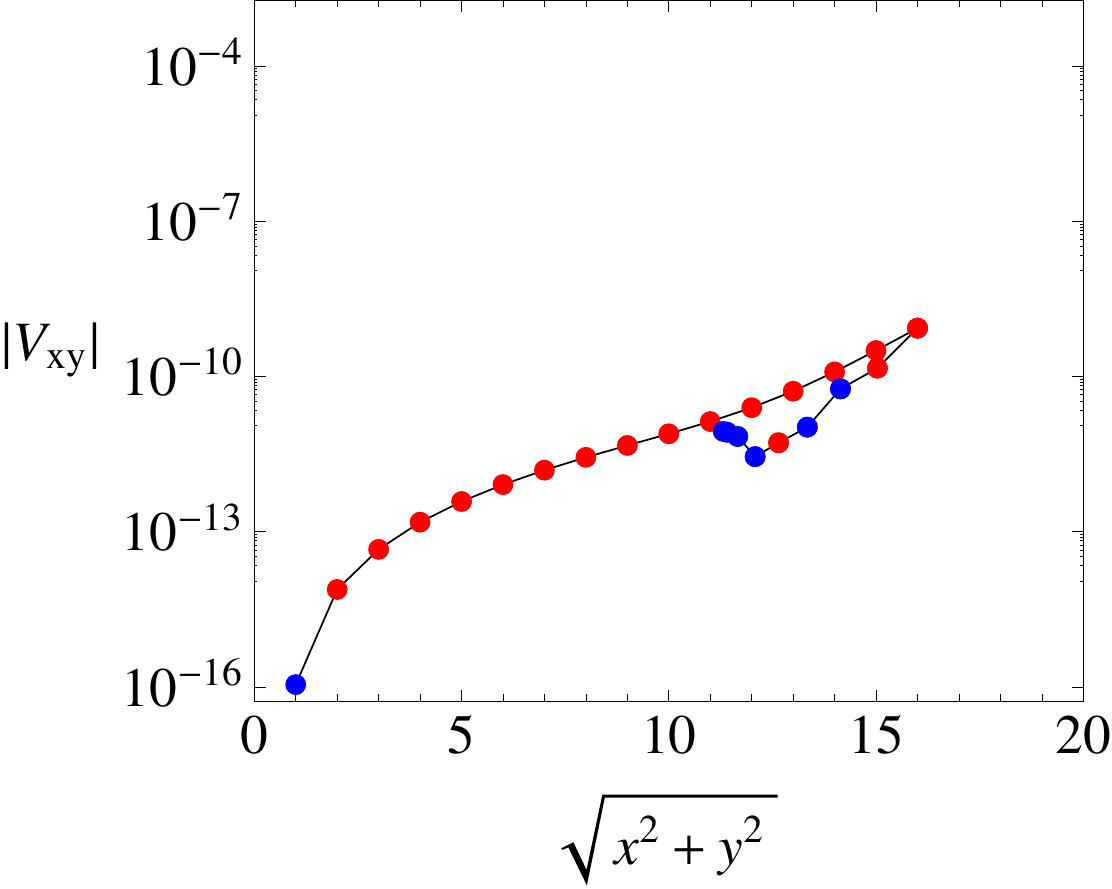}
		\label{AnnulusPotentialsAlongPath}
	}
	\caption{	
		(a) 
			Probability distribution $P^\text{circle}_{xy}$ 
				of a particle launched on a launching circle of radius $R_L = 20$
				with its position snapped to the nearest grid point.
			The electrostatic analogue is a discrete charge distribution $Q^\text{circle}_{xy}$.
			Note the pixelated appearance.
		(b)
			The resulting potential $V^\text{circle}_{xy}$ has fluctuations of order $\pm 10^{-4}$
				relative to $V_{00}$.  There is a strong hexadecapole ($\cos 4\phi$) component.
		(c)
			Potential along dashed-line path in previous panel.  
			Red (blue) indicate positive (negative) values of potential. 
			Potential fluctuations increase with distance from the center.
		(d) 
			Probability distribution $P^\text{annulus}_{xy}$ 
				of a particle launched on a fuzzy annulus 
				with inner and outer radii $R_{L1} = 16$ and $R_{L2} = 24$,
				where the radial distribution is governed by 
				a Kaiser-Bessel window function with parameter $\beta=15$.
		(e)
			The resulting potential $V_{xy}$ has much smaller fluctuations
				of order $\pm 10^{-10}$ (note difference in color scale).
		(f) 
			Potential along path.
	}
	\end{figure*}

We calculate the charges $Q_{x_0y_0}$ numerically according to Eq.~\eqref{eqSharpCircleProbabilities}, and we perform a fast 2D convolution with $F_{xy}$ to obtain $V_{xy}$.  Figures~\ref{CircleCharges}, \ref{CirclePotentials}, and \ref{CirclePotentialsAlongPath} show the charges and potentials for a launching circle of radius 20.  The ring of charge is distorted by snapping to the grid, leading to potential fluctuations on the scale of $10^{-4}$.

	%&&&&&&&&&&&&&&&&&&&&&&&&&&&&&&&&&&&&&&&&&&&&&&&&&&&&&&&&&&&&&&&&&&&&&&&&&&&&
	% FIGURE
	%&&&&&&&&&&&&&&&&&&&&&&&&&&&&&&&&&&&&&&&&&&&&&&&&&&&&&&&&&&&&&&&&&&&&&&&&&&&&

%----------------------------------------------------------------------------
\myheading{Eliminating bias using a fuzzy launching annulus:}
%----------------------------------------------------------------------------
It is useful to take some insights from computer graphics.  Early raster displays rendered oblique lines and circles with jagged edges.  Modern displays eliminate this problem using
antialiasing.\cite{freeman1974}  In the context of DLA, one might hope that launching bias might be reduced by ``antialiasing'' the launching circle.  The launching bias comes from the high-order Fourier components in $P(x,y)=\delta(\sqrt{x^2+y^2}-R_L)$ that cannot be represented on the grid.  Perhaps if we thicken the launching circle into an annulus and smear out its inner and outer boundaries, the resulting probability distribution will be smoother, and quantization error will be reduced.  We will show that this is indeed true.

Suppose we pick a radius from the probability distribution
	\begin{align}
	P_\text{rad} (r)
	&=
		\frac{2}{ R_{L2} - R_{L1} }
		P_\text{Kaiser} \Big(  
		\frac{ 2r - R_{L1} - R_{L2} }{R_{L2} - R_{L1}} 
	\Big)
	\end{align}
where	
	\begin{align}
	P_\text{Kaiser} (x) 
	&= \frac{ 
				\Theta(1-x^2)~
				I_0 \big( \beta \sqrt{1-x^2} \big) 
			}
				{
					\int_{-1}^1 dy~ I_0 \big( \beta \sqrt{1-y^2} \big)
				}
	\end{align}
where $I_0$ is the Bessel $I$ function.  This distribution corresponds to to a normalized Kaiser-Bessel window function\cite{kaiser1980} on the interval $[R_{L1},R_{L2}]$.  We choose the Kaiser-Bessel window because it has very small spectral leakage beyond the central lobe, and because the distribution is easy to evaluate and sample compared to the optimal Dolph-Chebyshev window.\cite{dolph1946}  We also pick an angle $\phi$ from the uniform distribution on $[0,2\pi)$, and set $x=\text{round} (R_L \cos \phi)$ and $y=\text{round} (R_L \sin \phi)$.  The probability distribution of the point $(x,y)$ is then
	\begin{align}
	P_{x_0y_0} &=
		\int_{x_0-\half}^{x_0+\half} dx~
		\int_{y_0-\half}^{y_0+\half} dy~
		\frac{
			P_\text{rad} \big( \sqrt{x^2+y^2} \big)  
		}{
			\sqrt{x^2+y^2}
		}
		.
	\end{align}
We calculate $Q$ and $V$ as before.  Figures~\ref{AnnulusCharges}, \ref{AnnulusPotentials}, and \ref{AnnulusPotentialsAlongPath} show the charges and potentials for a launching annulus with blurred edges ($R_{L1}=16, R_{L2}=24, \beta=15$).  We see that the potential in the interior of the annulus ($r<R_{L1}$) is uniform to within $10^{-10}$.
By increasing the inner radius and the thickness of the annulus, and by adjusting the parameter $\beta$, the potential can be made even more uniform.  For an annulus with $R_{L1}=40$, $R_{L2}=80$, and $\beta=24$, we find that $\abs{V_{xy}} < 10^{-14}$ for all $r<R_{L1}$.  For practical purposes this means that the launching bias has been eliminated.

In our DLA simulations we use a launching annulus with $R_{L1}=2R_B$, $R_{L2}=4R_B$, and $\beta=24$, where $R_B$ is the bounding radius of the cluster.  Since the cluster is well within the interior of the annulus, launching bias is negligible.  

We sample $x$ from $P_\text{Kaiser} (x)$ using rejection sampling\cite{mackayBook} with a Gaussian envelope function.  (In rejection sampling one must keep retrying until a move is accepted, unlike in the Metropolis algorithm.  A better term would be ``retrial sampling''.)

%============================================================================
\section{Eliminating killing bias\label{secKillingFree}}
%============================================================================
The second source of systematic error is the killing-and-relaunching.  We will eliminate this error by using an enormous killing circle of radius $R_K = 10^{14}$.

%============================================================================
\section{Accelerating diffusion outside the cluster using the walk-to-line algorithm}
%============================================================================
What is the catch in using large launching and killing circles?  Recall that a particle executing Brownian motion has a r.m.s. displacement that grows very slowly with time, $r_\text{rms} \propto \sqrt{t}$.  If a particle is launched $10^{9}$ sites away from the cluster, it will take at least $10^{18}$ timesteps before the particle has an appreciable probability of encountering the cluster!  Thus, in order to make a DLA simulation feasible, we must find a way to accelerate the diffusion process -- that is, we must ``fast forward'' through the random walk.  We do this using ``first passage theory.''

%----------------------------------------------------------------------------
\myheading{Electrostatic analogue of the first-passage problem:}
%----------------------------------------------------------------------------
Suppose a particle starts at $(x_0,y_0)$ and executes a random walk until it encounters a ``marked'' site $(x_n, y_n)$ where $n=1,2,3,\dotsc,N$.
We wish to find the probability distribution $p_{x_n y_n}$ of the final position of the particle.
This is known as the first-passage position, that is, the position at which an infinite random walk first passes through a marked site.
 
This Brownian problem maps to the following electrostatics problem.\cite{witten1983}   
Suppose a charge is placed at $(x_0,y_0)$ on a square lattice, and that the sites $(x_n, y_n)$ for $n=1,2,3,\dotsc,N$ are held at zero potential.
Solve the discrete Poisson equation:
	\begin{align}
	4V_{xy}  &{}   - V_{x+1,y} - V_{x-1,y} - V_{x,y+1} - V_{x,y-1} = Q_{xy},
		\nonumber\\
	V_{x_ny_n} &= 0,  \quad n=1,\dotsc,N
		\nonumber\\
	Q_{x_0y_0} &= 1,
		\nonumber\\
	Q_{x_ny_n} &= Q_n		\quad\text{(to be determined)},
		\nonumber\\
	Q_{xy} &= 0 \quad \text{on all other sites}
		.
	\end{align}
The charges $Q_{x_n y_n}$ give the desired first-passage probabilities.
In other words, we are to find the charge distribution induced on a grounded conducting object by an external point charge.
Rather than solving the discrete differential equation for the potential $V_{xy}$ in all space, it is better to solve the discrete integral equation for the charges $Q_{j}$ on the surface of the conductor.  This takes the form of $N$ simultaneous equations in $N$ variables.  Formally, we have
	\begin{align}
	V_i &= \sum_{j} G_{x_i-x_j, y_i-y_j} Q_j	+ G_{x_i-x_0, y_i-y_0} 
	\label{eqVGQ}
	\end{align}
where	 $i,j=1,2,\dotsc,N$ and $G_{xy}$ is the Green function of the square lattice Poisson equation such that
	\begin{align}
	4G_{xy} - G_{x+1,y} - G_{x-1,y} - G_{x,y+1} - G_{x,y-1} = \delta_x \delta_y 
	.
	\end{align}
The long-range logarithmic divergence in 2D poses two additional complications.  First, $G_{xy}$ contains an infinite constant.  We regularize this by defining the resistance Green function
$
	F_{xy} = G_{00} - G_{xy}
$.
As described in Appendix \ref{secGreenFunction}, $F_{xy}$ can be calculated to machine precision for any $x$ and $y$.  Take Eq.~\eqref{eqVGQ} and subtract the potential at a reference point $(x'',y'')$, which might as well be $(x_1We c,y_1)$.  This gives
	\begin{align}
	V_{i} - V_{x''y''} 
	&= \sum_{j} \big(  F_{x_i-x_j, y_i-y_j} - F_{x''-x_j, y''-y_j} \big) Q_j
			\nonumber\\&{}
		+ F_{x_i-x_0, y_i-y_0} -	F_{x''-x_0, y''-y_0} 
	.
	\label{eqVFQ}
	\end{align}
Second, in order for the potential to be well defined, the total charge in the entire system must be zero:
	\begin{align}
	\bigg( \sum_{j} Q_j  \bigg) + 1
		&=0
	.
	\label{eqTotalQ}
	\end{align}
Although Eq.~\eqref{eqVFQ} and \eqref{eqTotalQ} contain $N+1$ equations, they form a linear system of rank $N$, and so there is a unique solution for the $N$ charges, $Q_j$.  

	%&&&&&&&&&&&&&&&&&&&&&&&&&&&&&&&&&&&&&&&&&&&&&&&&&&&&&&&&&&&&&&&&&&&&&&&&&&&&
	% FIGURE
	%&&&&&&&&&&&&&&&&&&&&&&&&&&&&&&&&&&&&&&&&&&&&&&&&&&&&&&&&&&&&&&&&&&&&&&&&&&&&
	\begin{figure}[!htb]
	\subfigure[]{
		\includegraphics[width=0.45\columnwidth]{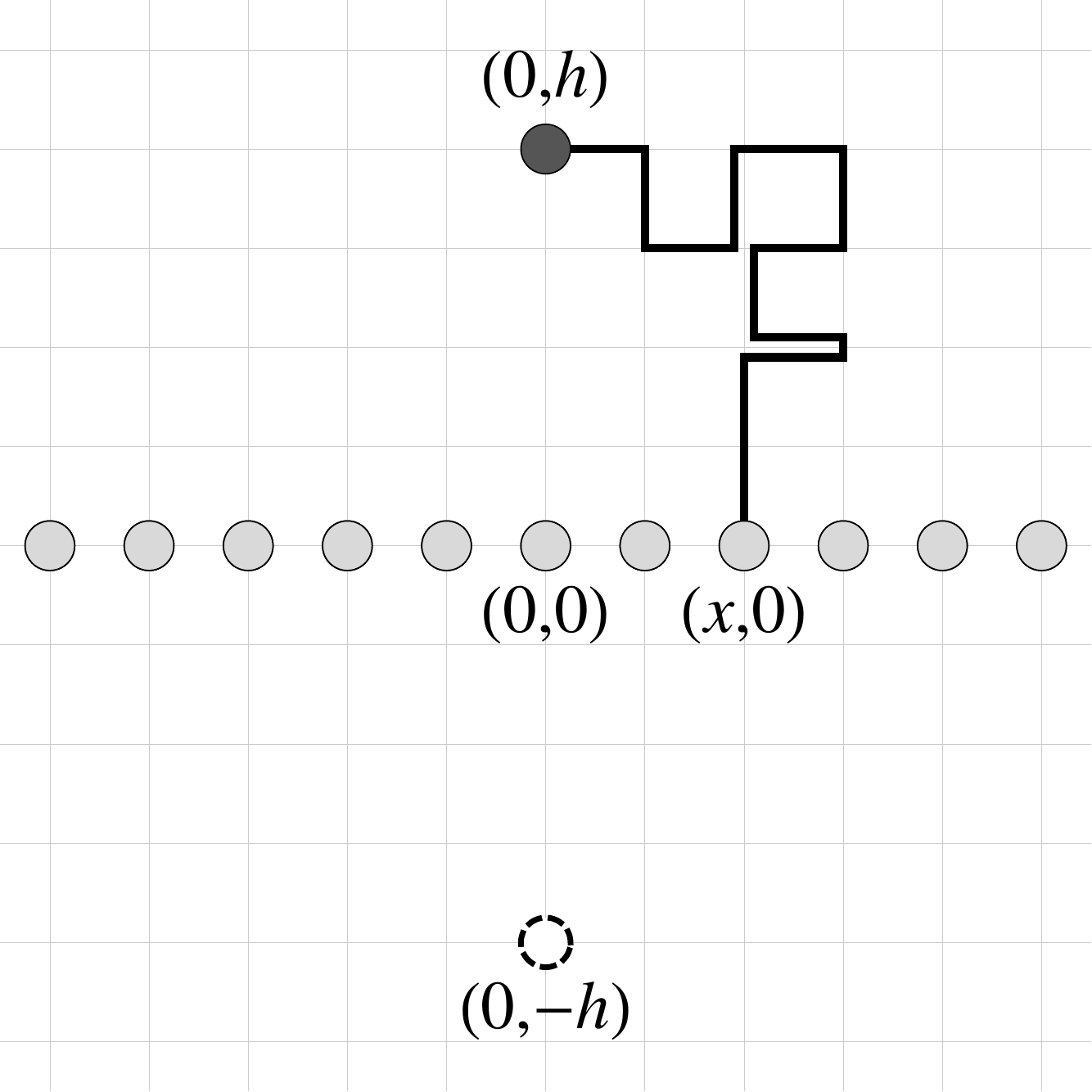}
		\label{WalkToLine}
	}
	\subfigure[]{
		\includegraphics[width=0.45\columnwidth]{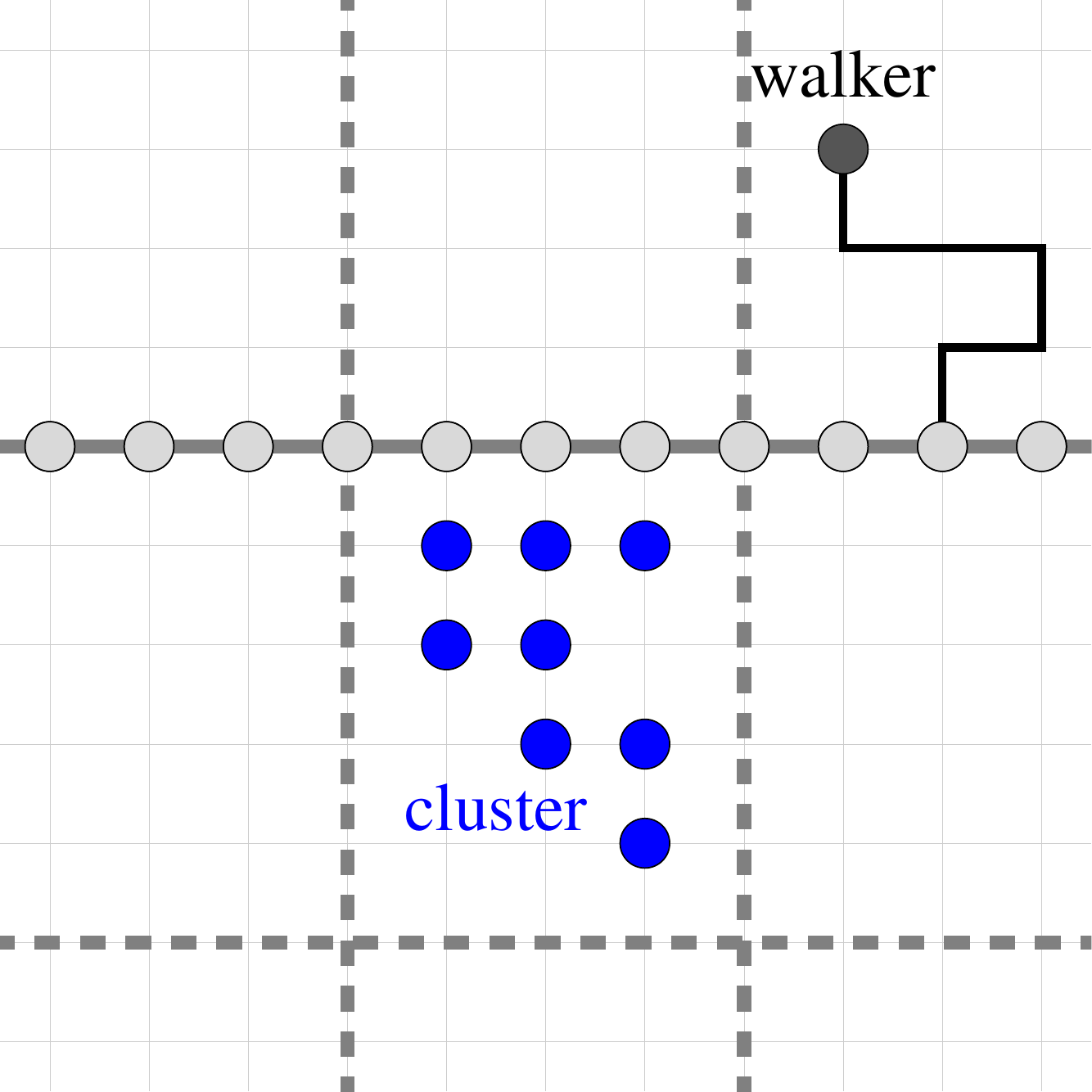}
		\label{WalkToBoundingBox}
	}
	\caption{	
		(a)
		Geometry for walk-to-line algorithm.  
		A particle (black disk) starts at $(0,h)$
			and executes a random walk on the square lattice.
		We wish to find the probability distribution $p_x$
			of the $x$-coordinate of first passage
			through the line $y=0$ (gray disks).
		The corresponding electrostatic problem can be solved
			using an image charge (dashed circle).
		(b)
		Application of walk-to-line algorithm to return the walker 
			towards the bounding box of the cluster.
	}
	\end{figure}

	%&&&&&&&&&&&&&&&&&&&&&&&&&&&&&&&&&&&&&&&&&&&&&&&&&&&&&&&&&&&&&&&&&&&&&&&&&&&&
	% FIGURE
	%&&&&&&&&&&&&&&&&&&&&&&&&&&&&&&&&&&&&&&&&&&&&&&&&&&&&&&&&&&&&&&&&&&&&&&&&&&&&
	\begin{figure}[!htb]
	\subfigure{
		\includegraphics[width=0.9\columnwidth]{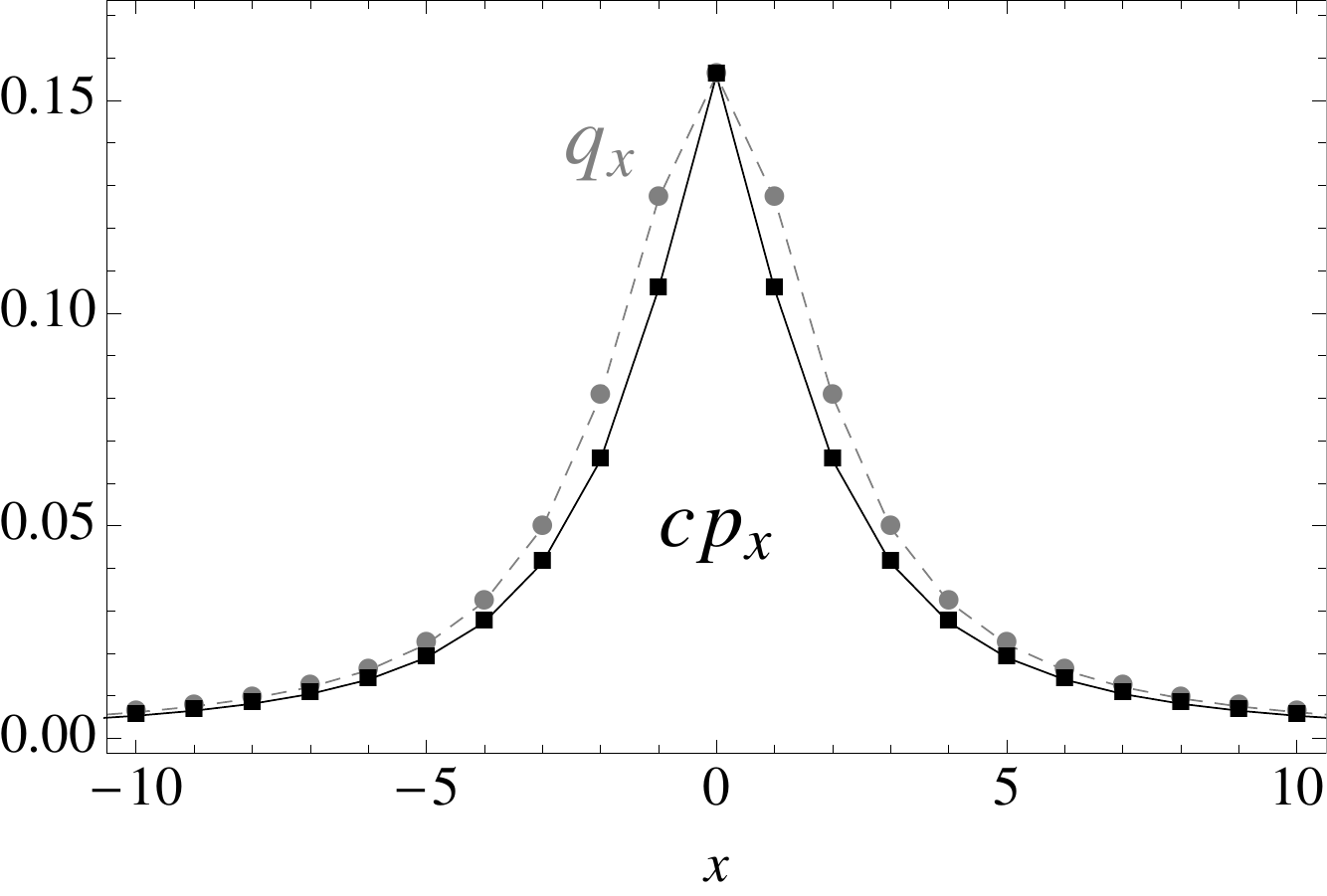}
		\label{RetrialSampling}
	}
	\caption{	
		Illustration of rejection sampling in the walk-to-line algorithm for $h=2$.  
		The gray symbols represent the envelope distribution $q_x$.
		The black symbols represent the target distribution $p_x$ scaled by a factor $c$,
		 chosen such $c p_x \leq q_x$ for all $x$.
	}
	\end{figure}

%----------------------------------------------------------------------------
\myheading{Walk-to-line algorithm:}
%----------------------------------------------------------------------------
We now apply this formalism to the situation shown in Fig.~\ref{WalkToLine}.  We will call this the walk-to-line (WtL) problem on a square lattice.  It is analogous to the walk-to-plane algorithm\cite{hwang2004,hwang2005} for 3D continuum Brownian diffusion.
\footnote{The original authors used the term ``walk-on-plane'' (WoP); here we write ``walk-to-plane'' and ``walk-to-line'' to avoid any misunderstanding.}

Suppose a charge is located at $(0,h)$ and every site on the $x$-axis $(x,0)$ is a conducting site held at zero potential.  The potential everywhere in the upper half-plane is given by the method of images as $V_{xy} = F_{x,y+h} - F_{x,y-h}$.  The electric field along the bond $(x,1)$--$(x,0)$ is $E_{x1,x0} = V_{x1} - V_{x0} = F_{x,h+1} - F_{x,h-1}$.  Mapping this back to the Brownian problem, we see that if a particle is launched at $(0,h)$, the probability distribution of first passage to the $x$-axis is 
	\begin{align}
	p_x &=  F_{x,h+1} - F_{x,h-1}.
	\end{align}
There are an infinite number of probabilities corresponding to all integer $x$.  We sample this distribution using rejection\footnote{Note that one must keep retrying until a move is accepted, as opposed to the Metropolis algorithm, where rejection is final.  The term \emph{rejection sampling} is misleading; a better term would be \emph{retrial sampling}.} sampling.\cite{mackayBook}  In order to use rejection sampling, we need an envelope function that resembles the target distribution (the probability distribution that we wish to sample from).  To obtain a suitable envelope function, consider the continuum version of the walk-to-line problem.  Suppose a particle starts at $(0,h)$ and executes continuum Brownian motion until it encounters the line $y=0$.  The electrostatic analog is a unit charge at $(0,h)$ and a grounded conducting plane at $y=0$.  Solve Poisson's equation with Dirichlet boundary conditions using the method of images.  The potential is $V(x,y)=\frac{1}{2\pi} \ln \sqrt{x^2 + (y+h)^2} - \frac{1}{2\pi} \ln \sqrt{x^2 + (y-h)^2}$, and so the charge density on the line is $\sigma(x)=-\frac{h}{\pi(x^2 + h^2)}$.  Thus the first-passage probability distribution is $P(x)=\frac{h}{\pi(x^2 + h^2)}$.  The cumulative distribution function (CDF) is $C(x)=\int_{-\infty}^x dx~P(x)=\frac{1}{\pi} \arctan \frac{x}{h} + \frac{1}{2}$.  Therefore, we can generate a sample from $P(x)$  using the inverse CDF method as $x = h \tan \pi (u - \half)$, where $u$ is a random number from the uniform distribution on $(0,1)$.  Now, since the target distribution is discrete, consider generating a sample using
	\begin{align}
	x = \text{round} \big(  h \tan \pi (u - \half)  \big)
	.
	\label{eqTangentRule}
	\end{align}
This corresponds to the discrete envelope distribution
	\begin{align}
	q_x &= \int_{x-\half}^{x+\half} dx'~ \frac{h}{\pi(x^2 + h^2)}
		\nonumber\\
	&= \frac{1}{\pi} \bigg( 
				\arctan \frac{x+\half}{h} 
			- \arctan \frac{x-\half}{h} 
		\bigg)
	%.
	\end{align}
The rejection sampling algorithm is as follows:
\begin{enumerate}
\item 
Find a number $c$ such that $cp_x \leq q_x$ for all $x$.  For our purposes we can choose $c=q_0/p_0$.
\item 
Draw a random integer $x$ from the envelope distribution $q_x$ (gray symbols in Fig.~\ref{RetrialSampling}).
\item 
Generate a uniform random number $v \in [0,1)$.  
\item 
If $v < c p_x/q_x$, return $x$.  Otherwise, go back to step 2.
\end{enumerate}
The returned value of $x$ is a sample from the target distribution $p_x$.

%----------------------------------------------------------------------------
\myheading{Application to square lattice DLA:}
%----------------------------------------------------------------------------
Suppose the diffusing particle is just outside the bounding rectangle of the cluster (see Fig.~\ref{WalkToBoundingBox}).  Apply the walk-to-line algorithm to return the particle to an infinite horizontal or vertical line parallel to the bounding rectangle.\cite{hwang2004,hwang2005}   A few iterations of this procedure usually suffice to return the particle to the bounding rectangle.

Now suppose the particle is a large distance $r$ away from the cluster.  Then, Eq.~\eqref{eqTangentRule} implies that the walk-to-line algorithm is roughly equivalent to multiplying $r$ by a random number $\alpha$ drawn from a Lorentzian distribution:
	\begin{align}
	r_\text{new} &= r \tan \pi(u-\half) = r \alpha
	,\quad P_\text{Lor} (\alpha) = \frac{1}{\pi(1+\alpha^2)}.
	\end{align}
Thus, the logarithm of the radius is incremented by a random number $\gamma=\ln\alpha$ drawn from a sech distribution:
	\begin{align}
	\ln \abs{r_\text{new}} &= \ln \abs{r} + \gamma
	,\quad P_\text{sech} (\gamma)=\frac{\sech \gamma}{\pi}.
	\end{align}
The variance of the sech distribution is 
	\begin{align}
	\int_{-\infty}^\infty d\gamma ~ \gamma^2 \frac{\sech \gamma}{\pi} 
	&= \frac{\pi^2}{4}.
	\end{align}
Therefore, $\ln\abs{r}$ executes a random walk with a step variance of $\pi^2/4$.  After $M=54$ iterations of the walk-to-line algorithm, the accumulated variance is $\frac{M \pi^2}{4}$, and one might expect $r$ to have increased or decreased by $\exp (\sqrt{M}{\pi/2}) \approx 10^5$.  If a walker somehow finds itself at a distance $10^{9}$ from a cluster of linear size $10^4$, after 54 iterations of walk-to-line, there is an appreciable probability that it will either have returned to the cluster (of radius $10^4$) or that it will have drifted outside the killing circle (of radius $10^{14}$).

In summary, if an errant particle finds itself a distance $10^{9}$ away from the cluster, the original Witten-Sander algorithm would take about $10^{18}$ steps to return it to the cluster, whereas the walk-to-line algorithm would take about $50$ iterations.  Although this is not perfect, it is certainly a great improvement.

%============================================================================
\section{Accelerating diffusion near the cluster using the walk-to-square algorithm}
\begin{figure}[!htb]
		\includegraphics[width=0.6\columnwidth]{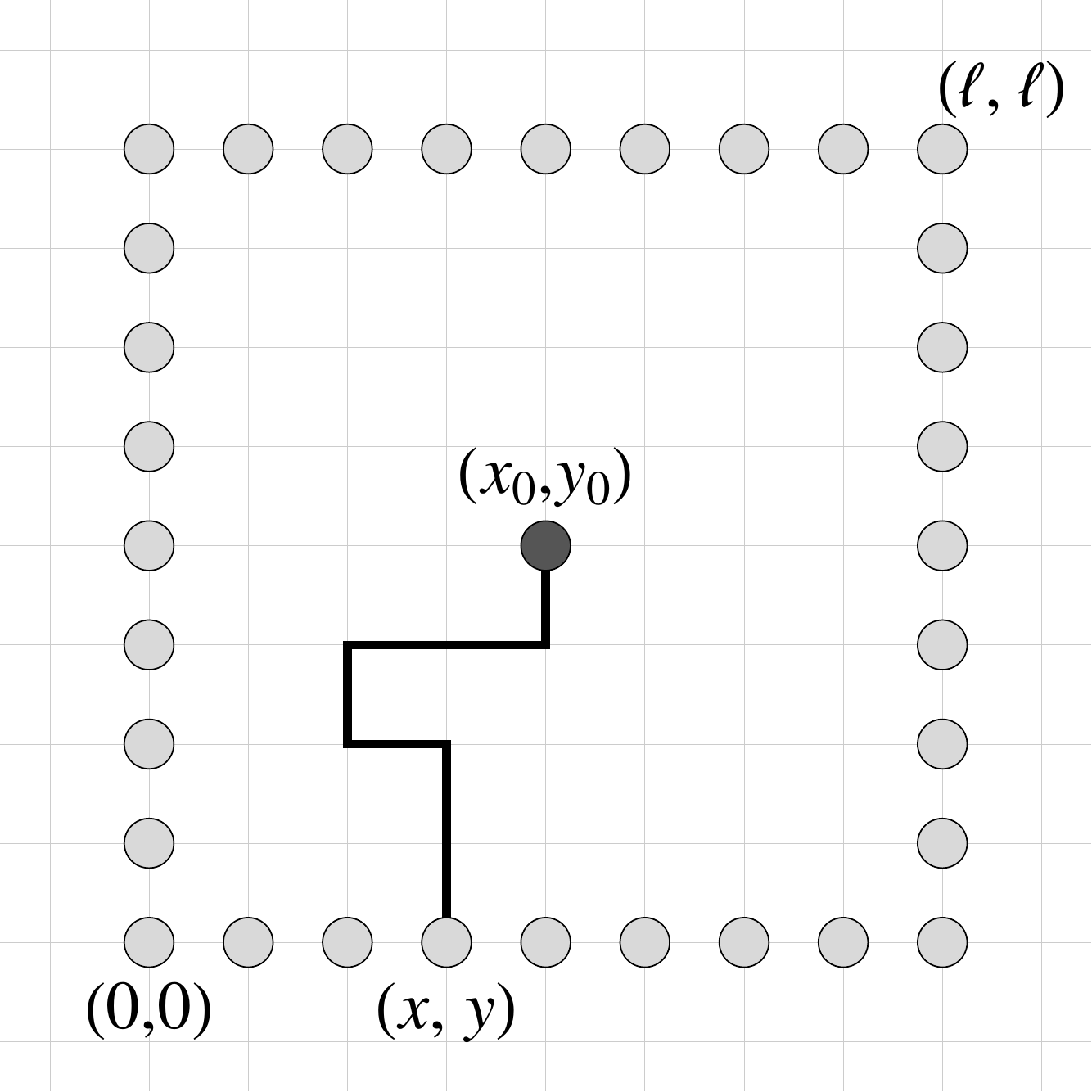}
	\caption{	
		\label{WalkOutToSquare}
		Geometry for walk-out-to-square algorithm.
		A particle starts at $(x_0,y_0)$ (black disk)
			and executes a random walk on the square lattice.
		We wish to find the probability distribution 
			of the first intersection of this walk
			with a surrounding square of side $l$ (gray disks).
	}
	\end{figure}

%============================================================================
Now let us consider another Brownian motion problem.  Suppose a random walker begins at point $(x_0,y_0)$ within a square with corners $(0,0)$ and $(l,l)$, as in Fig.~\ref{WalkOutToSquare}.  What is the probability that the walker makes first passage through the square at position $(x,y)$?

The corresponding electrostatic situation is a point charge at the center of a grounded conducting square.  We wish to solve the discrete Poisson equation with Dirichlet boundary conditions on a square:
	\begin{align}
	4V_{xy} &{} - V_{x+1,y} - V_{x-1,y} - V_{x,y+1} - V_{x,y-1} = Q_{xy},
		\nonumber\\
	V_{xy} &= 0  					~~~~~~~~~~~~~~~\text{$x_0\in\{0,l\}$ or $y_0\in\{0,l\}$},
		\nonumber\\
	Q_{xy} &= \delta_{x-x_0} \delta_{y-y_0}	~~~\text{$\{x,y\} \subseteq \{ 1,2,\dotsc,l-1 \} $}
		.
	\end{align}
This is a linear system involving a $(l-1)^2 \times (l-1)^2$ sparse matrix with integer coefficients.  Brute force matrix algebra gives the solutions $V_{xy}$ as rational numbers.  For example, for $l=8$ the solutions are
	\begin{align}
	\{V_{xy}\} &= \frac{1}{544}
	\begin{pmatrix}
	 	 9 & 18 & 26 & 30 & 26 & 18 & 9 \\
		18 & 37 & 56 & 68 & 56 & 37 & 18 \\
		26 & 56 & 93 & 130 & 93 & 56 & 26 \\
		30 & 68 & 130 & 266 & 130 & 68 & 30 \\
		26 & 56 & 93 & 130 & 93 & 56 & 26 \\
		18 & 37 & 56 & 68 & 56 & 37 & 18 \\
	 	 9 & 18 & 26 & 30 & 26 & 18 & 9 \\
	\end{pmatrix}
	\label{eqVxyForl8}
		.
	\end{align}
A better approach is to separate variables in Cartesians and superpose eigenfunctions to obtain the Green function.  Expand the charge distribution $Q_{xy}	= \delta_{x-x_0} \delta_{y-y_0}$ and the potential $V_{xy}$ in Fourier sine series, and connect them via the discrete Poisson equation:
	\begin{align}
	\widetilde{Q}_{pq}
	&= 
		\frac{2}{l}
		\sum_{x=1}^{l-1}
		\sum_{y=1}^{l-1} 
		\sin \frac{\pi p x}{l}
		\sin \frac{\pi q y}{l}
		Q_{xy}
		\label{eqQpq}
			\nonumber\\
	&=
		\frac{2}{l}
		\sin \frac{\pi p x_0}{l}
		\sin \frac{\pi q y_0}{l}
		,
			\\
	\widetilde{V}_{pq}
	&=
	\Big(
		4 - 2\cos \frac{\pi p}{l} - 2\cos \frac{\pi q}{l}
	\Big)^{-1} 	\widetilde{Q}_{pq} 
		,
		\label{eqVpq}
			\\
	V_{xy}
	&= 
		\frac{2}{l}
		\sum_{p=1}^{l-1} 
		\sum_{q=1}^{l-1} 
		\widetilde{V}_{pq}
		\sin \frac{\pi p x}{l}
		\sin \frac{\pi q y}{l}
		\label{eqDST}
		.
	\end{align}
We implement %Eqs.~\eqref{eqQpq}, \eqref{eqVpq}, and 
Eq.~\eqref{eqDST} using the fast 2D discrete sine transform (DST).  This allows us to compute $V_{xy}$ for all $x$ and $y$ in $O(l^2 \ln l)$ time.  This is faster than evaluating the double sums
	\begin{align}
	V_{xy}
	&= 
		\frac{4}{l^2}
		\sum_{p=1}^{l-1} 
		\sum_{q=1}^{l-1} 
		\frac{
		\sin \frac{\pi p x_0}{l}
		\sin \frac{\pi q y_0}{l}
		\sin \frac{\pi p x}{l}
		\sin \frac{\pi q y}{l}
		}{
				4 - 2\cos \frac{\pi p}{l} - 2\cos \frac{\pi q}{l}
		}
		,
	\end{align}
which takes $O(l^4)$ time.  For $l>512$ the DST method takes a long time and accumulates roundoff errors of the order of $10^{-12}$.  Then it becomes preferable to calculate $V_{xy}$ as a Madelung sum involving an infinite series of positive and negative image charges,
	\begin{align}
	V_{xy}
	&= 
		\sum_{m=-\infty}^{\infty}
		\sum_{n=-\infty}^{\infty}
		(-1)^{m+n}
		F_{x - (m+\half)l, y - (n+\half)l}
		~~ .
	\end{align}
For speed and accuracy, split $F_{xy}$ into a $\ln r$ part plus a correction due lattice anisotropy (Eq.~\eqref{eqGreenFunctionSeries}).  This gives $V_{xy}$ as the solution to the continuum problem (in terms of the Jacobi cn function) plus a lattice correction, which is best evaluated by grouping the charges into quadrupoles and truncating the sum appropriately.
 
Having found $V_{xy}$, we can find the electric field and hence the charge distribution on the boundary, $Q_{x0} = E_{x1,x0} = V_{x1} - V_{x0}$.  Thus the first-passage probabilities $p_{x0}$ are given simply by the first row of the $V_{xy}$ matrix, such as that in Eq.~\eqref{eqVxyForl8}.

	%&&&&&&&&&&&&&&&&&&&&&&&&&&&&&&&&&&&&&&&&&&&&&&&&&&&&&&&&&&&&&&&&&&&&&&&&&&&&
	% FIGURE
	%&&&&&&&&&&&&&&&&&&&&&&&&&&&&&&&&&&&&&&&&&&&&&&&&&&&&&&&&&&&&&&&&&&&&&&&&&&&&

\begin{figure}[!htb]
		\includegraphics[width=0.95\columnwidth]{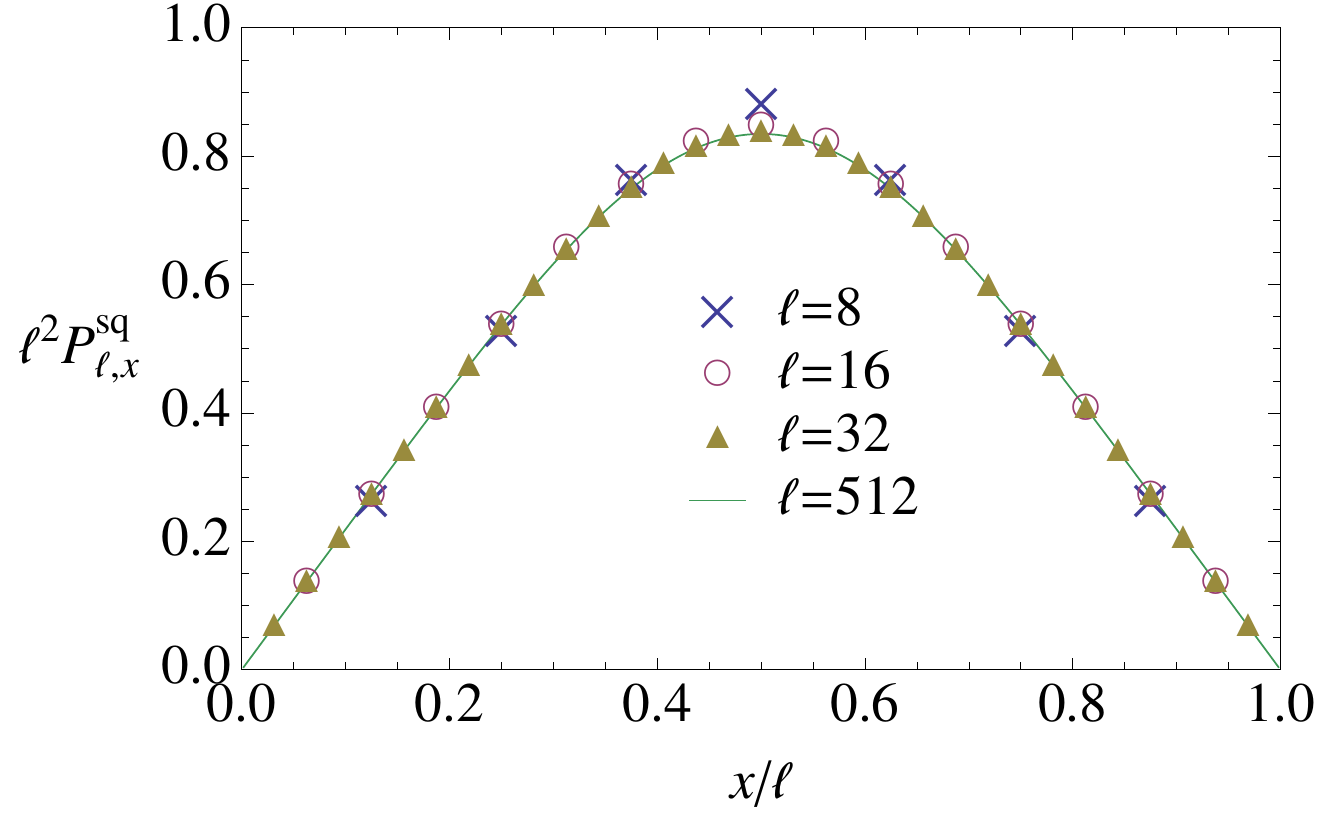}
	\caption{	
		\label{WalkOutToSquareProbs}
		First-passage probabilities $P^\text{sq}_{l,x}$ 
			from the center of a square of side $l$ 
			to every point $(x,0)$ on the lower edge,
			for $l=8, 16, 32, 512$.
		Horizontal and vertical axes are scaled to show
			that $l^2 P^\text{sq}$ approaches a universal function of $x/l$
			as $l\rightarrow \infty$.
		The universal function is the charge distribution 
			inside a grounded conducting square
			induced by a charge at its center.
	}
	\end{figure}

We have tabulated the first-passage probabilities $P^\text{sq}_{l,x}$ from the center of a square of side $l$ to every point $(x,0)$ on the lower edge, for $l \in \{ 2,4,8,16,32,64,128,256,512 \}$.  See Fig.~\ref{WalkOutToSquareProbs}.
These distributions can easily be sampled using precalculated Walker alias tables.\cite{walker1974,walker1977}

	%&&&&&&&&&&&&&&&&&&&&&&&&&&&&&&&&&&&&&&&&&&&&&&&&&&&&&&&&&&&&&&&&&&&&&&&&&&&&
	% FIGURE
	%&&&&&&&&&&&&&&&&&&&&&&&&&&&&&&&&&&&&&&&&&&&&&&&&&&&&&&&&&&&&&&&&&&&&&&&&&&&&

%----------------------------------------------------------------------------
\myheading{Application to square lattice DLA:}
%----------------------------------------------------------------------------
We store the DLA cluster as an array of 1's and 0's of dimensions $L \times L$, say.  We also maintain a hierarchy of coarse-grained representations of dimensions $\frac{L}{B} \times \frac{L}{B}$, representing $B\times B$ blocks, for block sizes $B=2,4,8,16,\dotsc,256$.  Every time a particle is added to the cluster, we mark the corresponding block at each level of the hierarchy as occupied.  The total amount of memory required is 
$L^2 + (\frac{L}{2})^2 + (\frac{L}{4})^2 + (\frac{L}{8})^2 + \dotso \approx \frac{4L^2}{3}$.  Thus, all the coarse arrays together require only $33\percent$ extra memory.  

\begin{figure*}[!tbh]
	\subfigure[]{
		\includegraphics[width=0.23\textwidth]{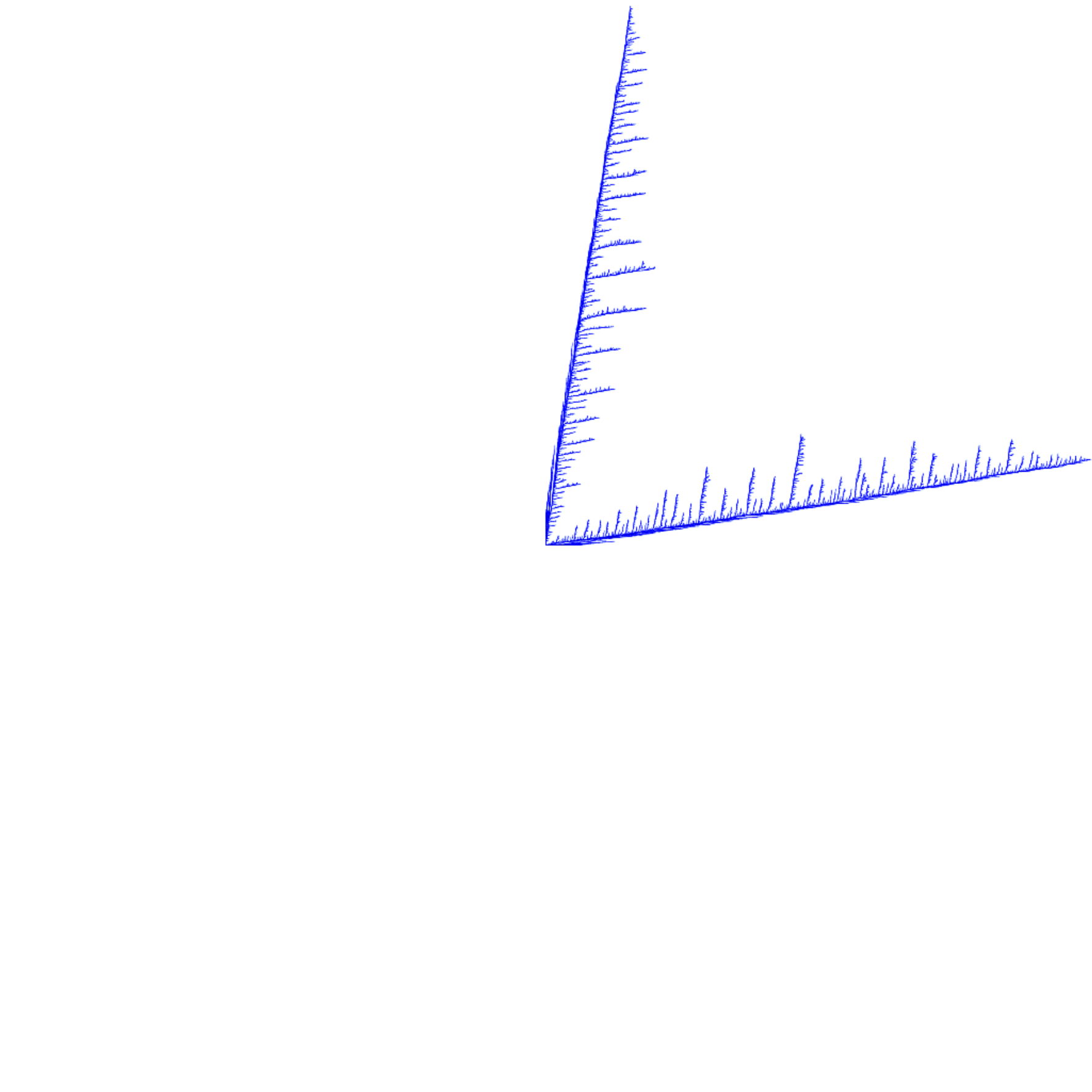}
		\label{Agg2-L65536-cluster}
	}
	\subfigure[]{
		\includegraphics[width=0.23\textwidth]{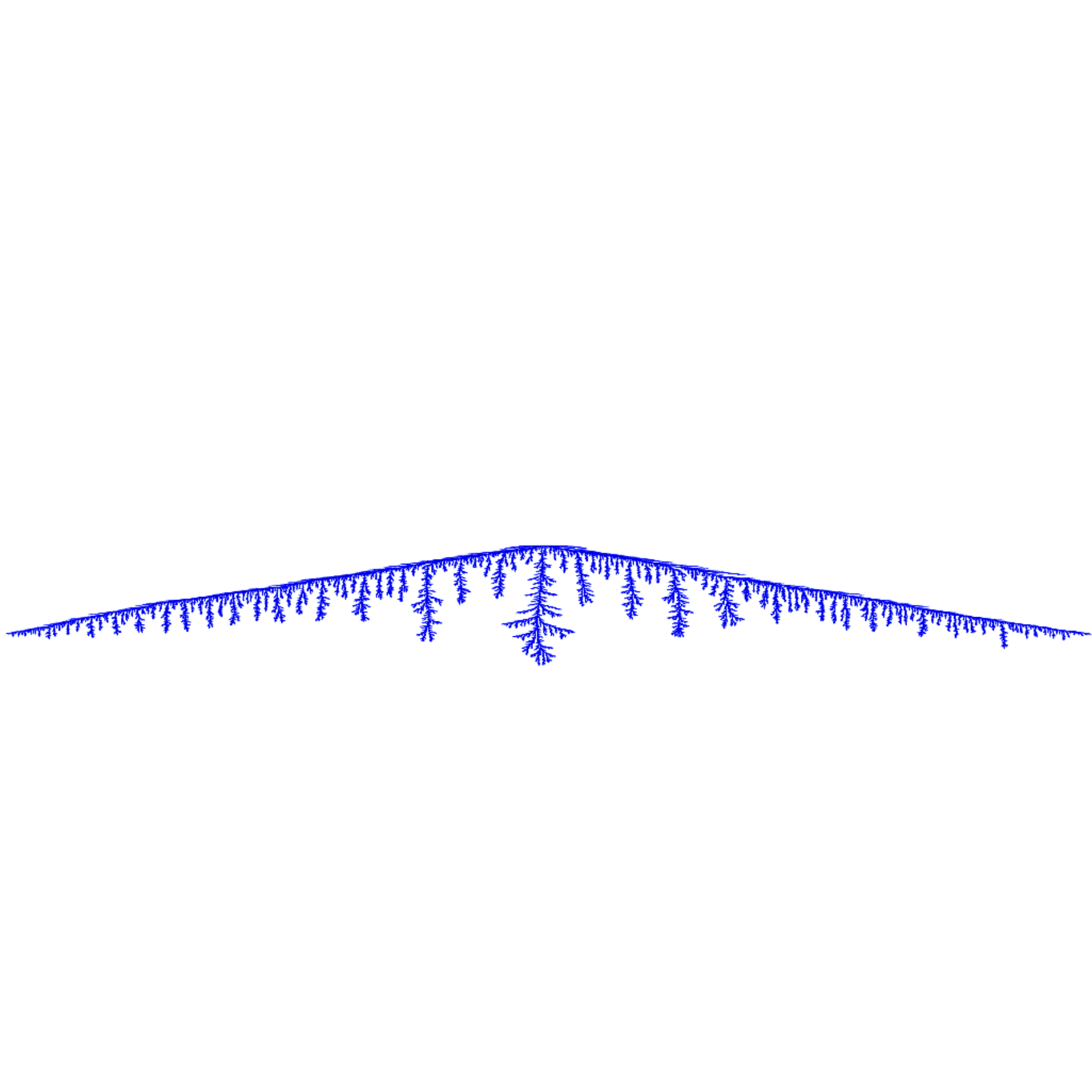}
		\label{Agg3-L65536-cluster}
	}
	\subfigure[]{
		\includegraphics[width=0.23\textwidth]{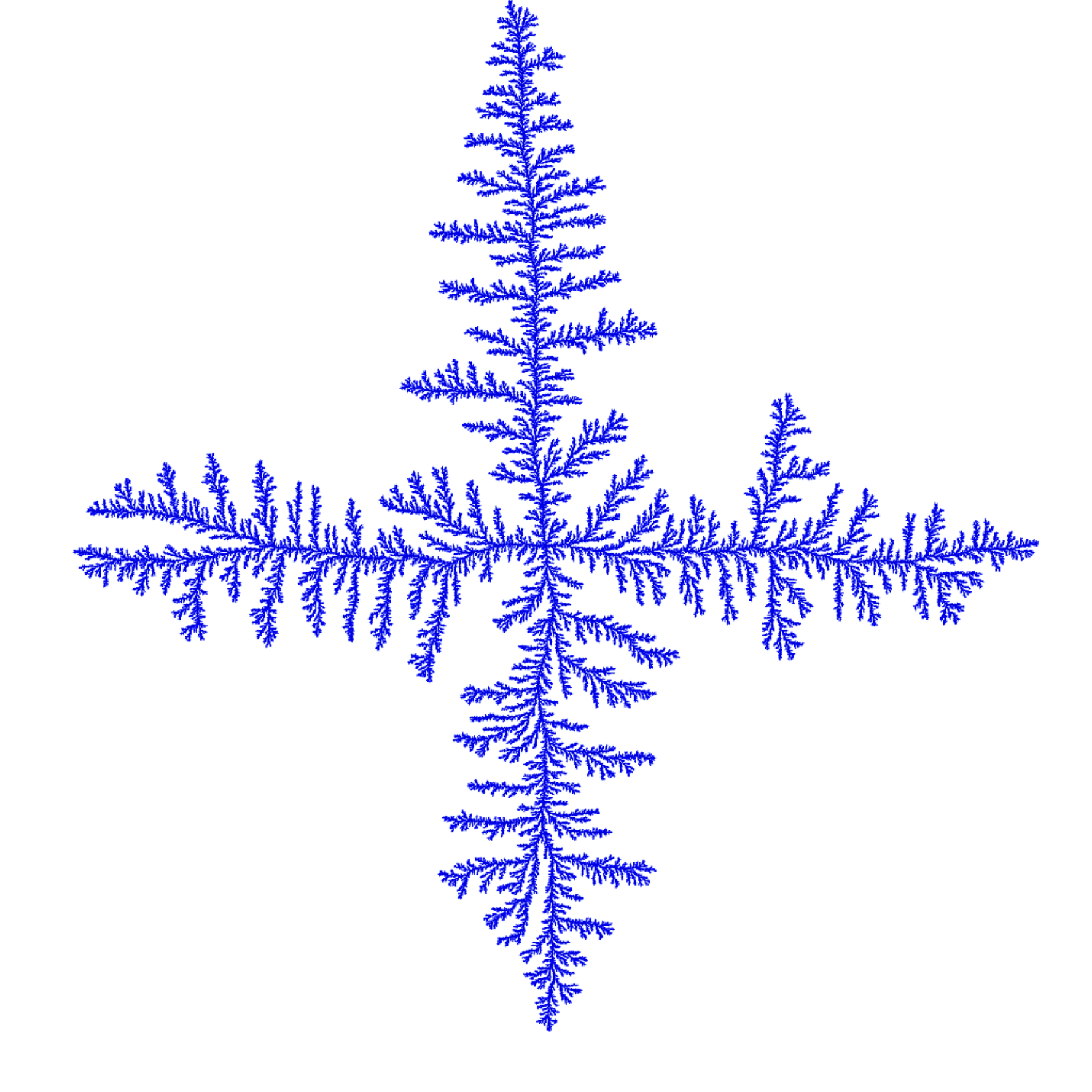}
		\label{Agg4-L65536-cluster}
	}
	\subfigure[]{
		\includegraphics[width=0.23\textwidth]{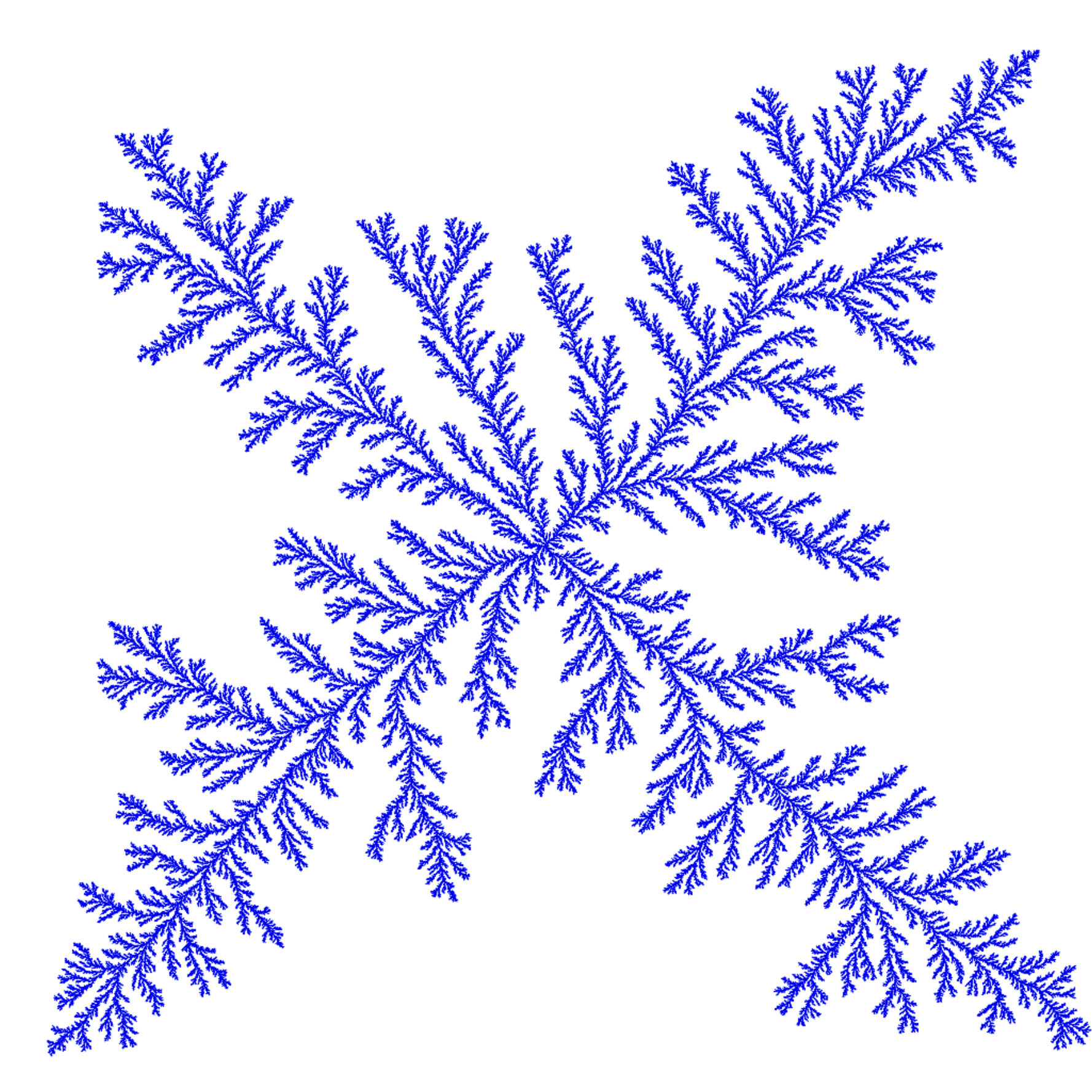}
		\label{Agg8-L65536-cluster}
	}
	\caption{	
		Anisotropic growth of lattice DLA clusters.
		In each case the seed cluster was a single particle.
		New particles from far away executed random walks
		 according to a 4-neighbor diffusion rule.
		Each particle became stuck to the cluster 
			according to 2-, 3-, 4-, or 8-neighbor aggregation rules.
		The simulation was terminated when the cluster reached
		 	the edge of a $65536\times 65536$ square lattice.
		The masses of the clusters were
			 6\,864\,668, 14\,638\,988, 96\,244\,639, and 191\,792\,092 respectively.
	}
	\end{figure*}

Suppose the diffusing particle is within the bounding box of the cluster, so that the walk-to-line algorithm is inapplicable.  We can revert to moving the particle one step at a time.  However, if the particle is in a cavity of radius $R_C$, it will take of the order of $(R_C)^2$ steps for the particle to find its way to an occupied site.  From Fig.~\ref{Agg4-L65536-cluster} we can see that a particle in the corner of the bounding box would spend $10^7$ timesteps wandering around in a cavity of linear size $4000$.  

Fortunately, the walk-to-square algorithm allows us to speed up the diffusion process.  We start at the coarsest level of the hierarchy ($256\times 256$ blocks) and examine the $3\times 3$ array of blocks around the diffusing particle's block.  If any one of these nine blocks is occupied, we proceed to the next finer level of the hierarchy.  We repeat this until we get to a level where all nine blocks are empty.  This means that we can move the particle by $\pm B$ units in the $x$ direction and $\pm B$ units in the $y$ direction without contacting the cluster.  Therefore we can apply the walk-to-square algorithm for a square of side $2B$ centered on the diffusing particle.  

Many authors have used hierarchical representations and variable stepsizes.\cite{meakin1983pra3,menshutin2006}  Meakin et al.\cite{meakin1987} used a combination of off-lattice jumps and on-lattice steps.  Ball et al.\cite{ball1985jpa} used a lookup table calculated using Laplace's equation to compute the first passage to the square in a manner accurate to better than $1\percent$.  However, to the best of our knowledge, our walk-to-square algorithm, which is based on exact Green functions, is the first \emph{unbiased} variable-stepsize algorithm for \emph{lattice} DLA.

%============================================================================
\section{Efficient bias-free algorithm for DLA}
%============================================================================
Having described all the ingredients, we now give a summary of our DLA simulation algorithm, omitting optimization details:

\begin{enumerate}
\item 
Set up a seed cluster.
\item 
Launch a new walker on a fuzzy annulus of inner radius $R_{L1} = 2R_B$ and outer radius  $R_{L1} = 4R_B$, where $R_B$ is the bounding radius of the cluster (see Fig.~\ref{AnnulusCharges}).
\item 
If the walker lies outside the bounding rectangle of the cluster, use the walk-to-line algorithm to move the walker to an infinite line on the near side of the bounding box (see Fig.~\ref{WalkToLine}).
If this takes the walker outside the killing circle of radius $R_K=10^{14}$, discard the walker and go back to step 2.
\\~ \\
If instead the walker lies inside the bounding rectangle, start at the coarsest level of the hierarchy, and proceed to finer and finer levels until one finds a scale at which all eight neighboring blocks contain no sticky sites.  Then, apply the walk-out-to-square algorithm to move the walker to an edge of a square contained within the empty region (see Fig.~\ref{WalkOutToSquare}).
\item
If the walker is now at a sticky site, freeze it (i.e., add it to the cluster and mark its neighbors as new sticky sites) and go to step 2.  Otherwise, go to step 3. 
\end{enumerate}

In this algorithm all probability distributions are sampled with errors less than $10^{-12}$.  Thus systematic error is practically eliminated, leaving statistical error and finite cluster size as the only sources of error.

\begin{figure}[!htb]
	\subfigure[]{
		\includegraphics[width=.15\columnwidth]{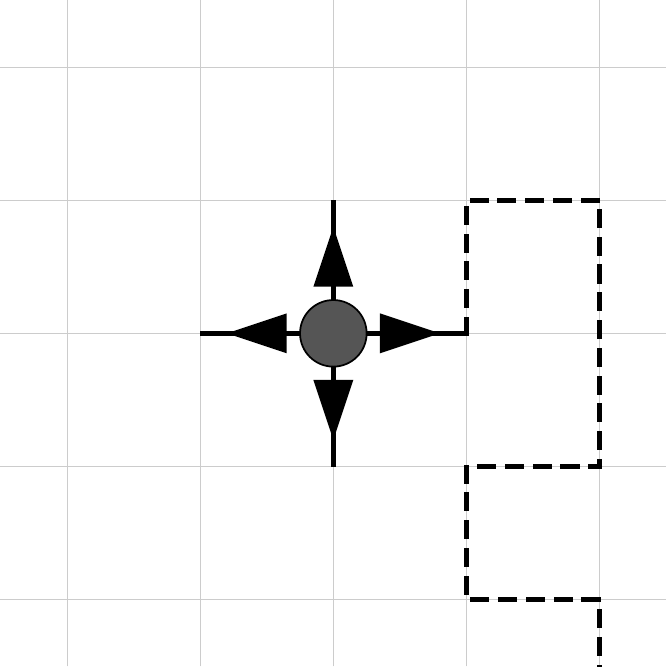}
		\label{RuleForDiffusion}
	}
	\subfigure[]{
		\includegraphics[width=.125\columnwidth]{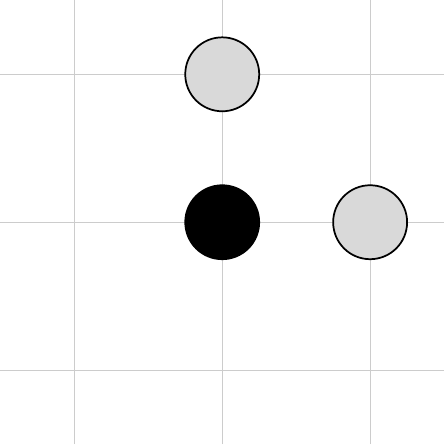}
		\label{Agg2}
	}
	\subfigure[]{
		\includegraphics[width=.125\columnwidth]{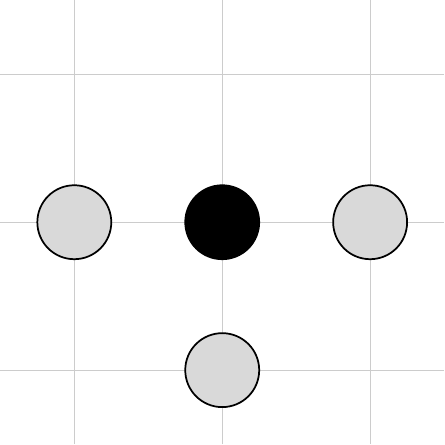}
		\label{Agg3}
	}
	\subfigure[]{
		\includegraphics[width=.125\columnwidth]{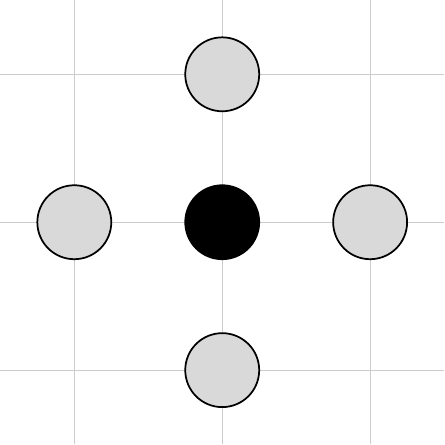}
		\label{Agg4}
	}
	\subfigure[]{
		\includegraphics[width=.125\columnwidth]{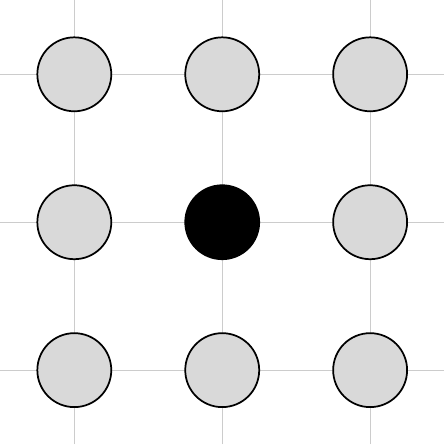}
		\label{Agg8}
	}
	\caption{
		(a) The diffusing particle moves to one of 4 neighboring sites
			with equal probability.
		(b,c,d,e): When the particle freezes at a sticky site (black),
			2, 3, 4, or 8 neighboring sites (gray) are marked as sticky sites
			according to the aggregation rule.
	}
	\end{figure}

We have implemented this algorithm in C++ (see Supplementary Material).  We use 64-bit floating-point arithmetic, and we represent $(x,y)$ coordinates by 64-bit integers to allow the killing circle radius to be $R_K=10^{14}$.
We assume nearest-neighbor diffusion, such that the diffusing particle only moves horizontally or vertically (Fig.~\ref{RuleForDiffusion}).
We try various aggregation rules (Figs.~\ref{Agg2}, \ref{Agg3}, \ref{Agg4}, and \ref{Agg8});
square lattice DLA studies in the literature typically use the 4-neighbor rule, where sticky sites are horizontally or vertically adjacent to a cluster site.
At each level of the hierarchy we store the pattern of frozen/sticky sites using a bit array.  For simplicity we do not distinguish frozen sites from sticky sites.

	%&&&&&&&&&&&&&&&&&&&&&&&&&&&&&&&&&&&&&&&&&&&&&&&&&&&&&&&&&&&&&&&&&&&&&&&&&&&&
	% FIGURE
	%&&&&&&&&&&&&&&&&&&&&&&&&&&&&&&&&&&&&&&&&&&&&&&&&&&&&&&&&&&&&&&&&&&&&&&&&&&&&

In this work we have used a built-in system random number generator, which is a non-linear additive feedback generator with a period of approximately $16(2^{31} - 1)$.  We have not tested the effect of different random number generators.

	%&&&&&&&&&&&&&&&&&&&&&&&&&&&&&&&&&&&&&&&&&&&&&&&&&&&&&&&&&&&&&&&&&&&&&&&&&&&&
	% FIGURE
	%&&&&&&&&&&&&&&&&&&&&&&&&&&&&&&&&&&&&&&&&&&&&&&&&&&&&&&&&&&&&&&&&&&&&&&&&&&&&

	%&&&&&&&&&&&&&&&&&&&&&&&&&&&&&&&&&&&&&&&&&&&&&&&&&&&&&&&&&&&&&&&&&&&&&&&&&&&&
	% FIGURE
	%&&&&&&&&&&&&&&&&&&&&&&&&&&&&&&&&&&&&&&&&&&&&&&&&&&&&&&&&&&&&&&&&&&&&&&&&&&&&
	\begin{figure}[!htb]
	\subfigure{
		\includegraphics[width=0.98\columnwidth]{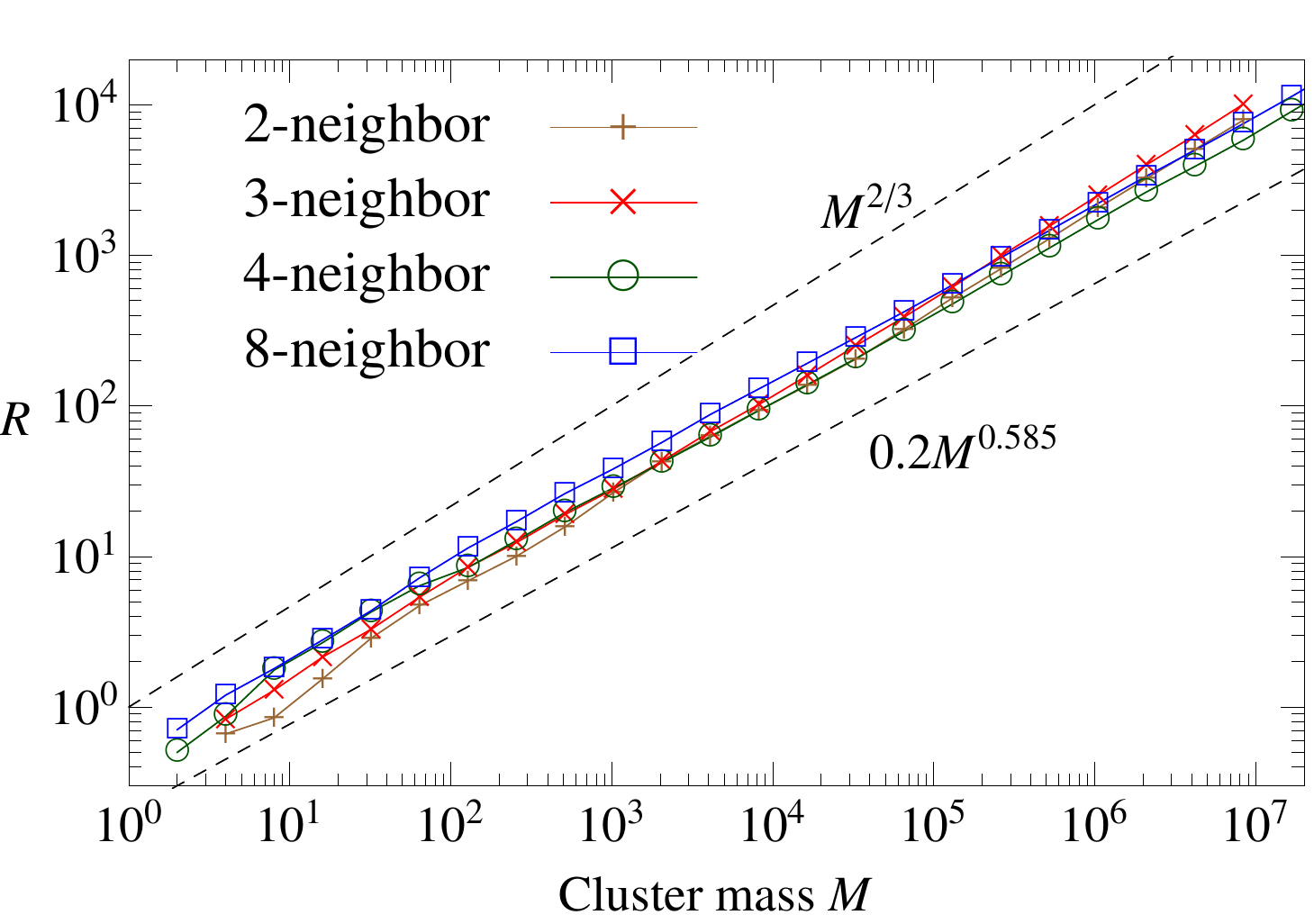}
		\label{RvsM}
	}
	\subfigure{
		\includegraphics[width=0.98\columnwidth]{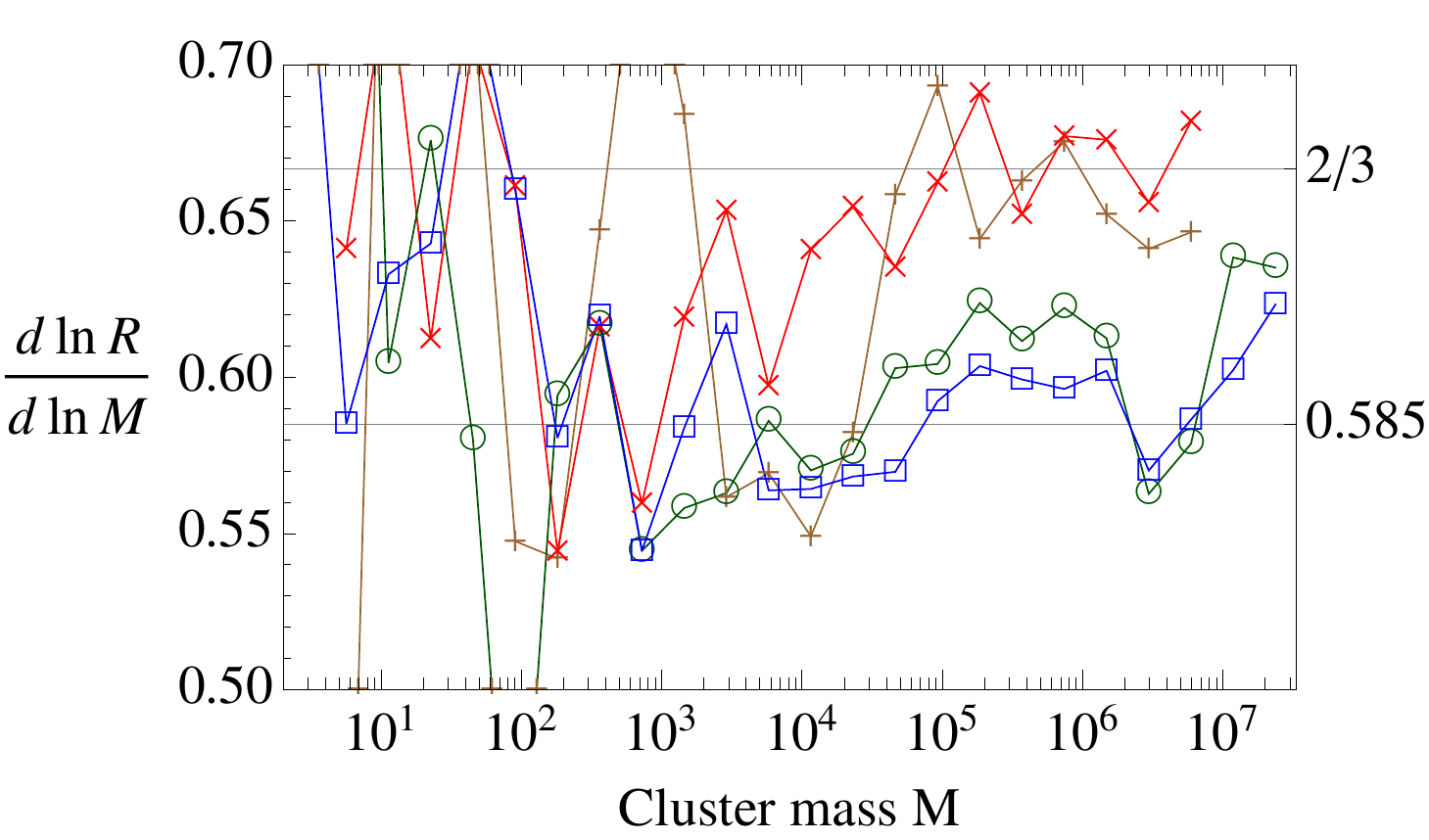}
		\label{dlnRdlnM}
	}	
	\caption{	
		\label{RadiusVsMass}
		(Top) 
			Radius of gyration $R$ as a function of cluster mass $M$
			for 2-, 3-, 4-, and 8-neighbor aggregation rules,
			 during the growth of single clusters.
			The differences are difficult to distinguish on this type of plot.
		(Bottom) 
			Estimates of the exponent using $\beta(M)=d(\ln R)/d(\ln M)$.
	}
	\end{figure}

%============================================================================
\section{Results\label{secResults}}
%============================================================================
%----------------------------------------------------------------------------
\myheading{Cluster shape:}
%----------------------------------------------------------------------------
Figures~\ref{Agg2-L65536-cluster}, \ref{Agg3-L65536-cluster}, \ref{Agg4-L65536-cluster}, and \ref{Agg8-L65536-cluster} show square lattice DLA clusters grown using different aggregation rules.
The 2-neighbor and 3-neighbor aggregation rules clearly manifest themselves by producing $L$-shaped and $T$-shaped clusters.
The 4-neighbor aggregation rule produces faster growth along horizontal and vertical directions, whereas the 8-neighbor aggregation rule produces faster growth along diagonal directions; both these rules lead to a $\cos 4\phi$ asymmetry in the angular mass distribution.

We have tried starting with seed clusters of various shapes ($\circ$, $+$, $\times$, $-$, $/$).  Regardless of the shape of the seed cluster, the growing cluster evolves toward a shape  determined by the aggregation rule.
Thus the asymptotic shape of the cluster is governed by the aggregation rule (the way in which particles stick together),\cite{ball1985jpa}   and not by the diffusion rule, nor by the seed cluster shape.

Various authors have reported that 
a ``square deposition habit'' leads to $\diamond$ and $+$ cluster shapes,\cite{ball1985jpa,meakin1987} 
whereas a ``diagonal deposition habit'' leads to $\times$ shapes.\cite{ball1985jpa}   
For off-lattice DLA, it was found that the ratio of the principal radii of gyration tends to unity for large clusters\cite{garik1985} -- i.e., if a cluster happens to start off with an elliptical shape, it evolves towards a circular shape.
Our results agree with these statements.

%----------------------------------------------------------------------------
\myheading{Fractal dimension:}
%----------------------------------------------------------------------------
Figure~\ref{RvsM} shows the radius of gyration $R$ as a function of the cluster mass $M$, during the growth of a single cluster, for various aggregation rules.  The data are roughly consistent with a power law $R\propto M^\beta$ somewhere between $\beta=0.585=1/1.71$, which is the exponent for 2D continuum DLA,\cite{meakin1985} and $\beta=0.667=2/3$, which is Kesten's upper bound for 2D DLA.
Figure~\ref{dlnRdlnM} shows $\beta(M) = d(\ln R)/d(\ln M)$ estimated from ratios between successive $(R,M)$ data points.
For 2- and 3-neighbor aggregation rules, $\beta(M)$ is close to $2/3$ for large $M$.
For 4- and 8-neighbor aggregation rules $\beta(M)$ appears to be close to $0.585$ for moderate $M$, but for very large $M$ it appears possible that $\beta(M)$ is increasing towards $2/3$.

Meakin et al.\cite{meakin1987} reported that $\beta$ evolves from $0.585$ towards $0.667 = 2/3$ during the growth of the cluster for square lattice DLA.
Menshutin et al.\cite{menshutin2011} also reported that $\beta\rightarrow 2/3$ for a variant of DLA in which particles diffuse via continuum 2D Brownian motion but aggregate onto lattice sites using an ``antenna'' rule.
Our results are consistent with these statements.

%============================================================================
\section{Conclusions\label{secConclusions}}
%============================================================================
We present an improved algorithm for 2D lattice DLA that reduces systematic errors in probabilistic sampling below $10^{-12}$.
We build clusters of $10^8$ particles on lattices of size $65536\times 65536$.
We verify that the anisotropy of the aggregation process leads to anisotropy of the cluster shape, so that the radius-of-gyration exponent evolves from $\beta=0.585=1/1.71$ towards $\beta=0.667=2/3$.

Our unbiased DLA algorithm can be generalized to triangular lattices, cubic lattices, and other lattices.  There are analytic expressions for triangular lattice Green functions,\cite{kleinertBook,atkinson1999} and Green functions on 3D lattices can be calculated numerically.\cite{cserti2000}

We are grateful to William Schwalm for helpful discussions.

%COMPARISON WITH LITERATURE:
%ball1985jpa (Ball, Brady, Rossi, and Thompson)
%
%meakin1983pra2, meakin1983pra3 
%Nearest-neighbor.  (EWNS deposition habit)
%
%meakin1987 (Meakin, Brady, Ramanlal, and Sander)
%Same algorithm as ball1985jpa.
%Varying the sizes of the launching circle and killing circle did not introduce measurable distortion.
%Radius of gyration vs cluster mass:
%For small clsuters beta=0.585=1/1.710 (unuivrsal) but for large clusters beta=0.61=1/1.64.
%SHAPE: +

%(Note: Menshutin shows reuslts for $5\times 10^7$ particles; and even claims to be able to do $2\times 10^8$ particles with swapping)

%\newpage
\appendix
%============================================================================
\section{Square lattice resistance Green function\label{secGreenFunction}}
%============================================================================
In this appendix we consider the Green function
	\begin{align}
	G_{xy} &=
	 \int_0^{2\pi} \frac{dp}{2\pi}
	 \int_0^{2\pi} \frac{dq}{2\pi}
	 ~
	 \frac{e^{ipx+iqy}}{4 - e^{ip} - e^{-ip} - e^{iq} - e^{-iq} }
	.
	\label{eqGIntegral}
	\end{align}
and the regularized Green function
	\begin{align}
	F_{xy} &=
	 \int_0^{2\pi} \frac{dp}{2\pi}
	 \int_0^{2\pi} \frac{dq}{2\pi}
	 ~
	 \frac{1 - e^{ipx+iqy} }{4 - e^{ip} - e^{-ip} - e^{iq} - e^{-iq} }
	.
	\label{eqFIntegral}
	\end{align}

%----------------------------------------------------------------------------
\myheading{Recursion relations:}
%----------------------------------------------------------------------------
By symmetry it can be shown that $F_{00}=0$ and $F_{10}=\frac{1}{4}$.  Using complex variable techniques it can be shown that\cite{kleinertBook}
	\begin{align}
	F_{xx} &= \frac{1}{\pi} \sum_{n=1}^x \frac{1}{2n-1} = \frac{2H_{2x} - H_x}{2\pi}
	\label{eqHarmonicNumbers}
	\end{align}
where $H_x=1+\frac{1}{2}+\frac{1}{3}+\dotso+\frac{1}{x}$ are the harmonic numbers.  $F_{xy}$ also satisfies the discrete Poisson equation
	\begin{align}
	4F_{xy} - F_{x+1,y} - F_{x-1,y} - F_{x,y+1} - F_{x,y-1} = - \delta_x \delta_y 
	.
	\label{eqRecursion}
	\end{align}
In principle, Eqs.~\eqref{eqHarmonicNumbers} and \eqref{eqRecursion} allow one to compute $F_{xy}$ for all $x$ and $y$.  However, this procedure is unstable to roundoff error if implemented numerically.  Thus, we implement the recursion relations using symbolic algebra, and use extra-precision arithmetic to convert the results to floating-point numbers.  The first few $F_{xy}$ are shown in Table~\ref{tabResistanceGreenFunction}.\cite{kleinertBook,atkinson1999,cserti2000}

	\begin{table}[!htb]
	\begin{align}
	%\begin{ruledtabular}
	\begin{array}{c|ccccc}
		%\hline \vspace{-4mm}\\ 
		\hline
	F_{xy}
	  & 0 & 1 & 2 & 3 \\
	 \hline
	0 & 0 & \frac{1}{4} & 1 - \frac{2}{\pi} & \frac{17}{4} - \frac{12}{\pi} \\
	1 & \frac{1}{4} & \frac{1}{\pi} & -\frac{1}{4} + \frac{2}{\pi} & -2 + \frac{23}{3\pi} \\
	2 & 1 - \frac{2}{\pi} &  -\frac{1}{4} + \frac{2}{\pi} & \frac{4}{3\pi} & \frac{1}{4} + \frac{2}{3\pi} \\
	3 &  \frac{17}{4} - \frac{12}{\pi} & -2 + \frac{23}{3\pi} & \frac{1}{4} + \frac{2}{3\pi} & \frac{23}{15\pi} \\
		\hline
	\end{array}
	%\end{ruledtabular}
	\end{align}
	\caption{
		Values of the square lattice resistance Green function.
		\label{tabResistanceGreenFunction}
	}
	\end{table}

%----------------------------------------------------------------------------
\myheading{Series approximation at large distances:}
%----------------------------------------------------------------------------
For large values of $x$ and $y$, the behavior of Eq.~\eqref{eqGIntegral} is dominated by small $p$ and $q$.  Expand the denominator in powers of $p=k\cos\phi$ and $q=k\cos\phi$:
	\begin{align}
	& 4 - e^{ip} - e^{-ip} - e^{iq} - e^{-iq}
		\nonumber\\
	&= 4 - 2 \cos p - 2 \cos q
		\nonumber\\
	&= k^2 
		- \tfrac{k^4}{12} (\cos^4\phi + \sin^4\phi) 
		+ \tfrac{k^6}{360} (\cos^6\phi + \sin^6\phi) 
		- \dotso
	.
	\end{align}
Expand the reciprocal in powers of $k$:
	\begin{align}
	& \frac{1}{4 - e^{ip} - e^{-ip} - e^{iq} - e^{-iq}}
		\nonumber\\
	&=\tfrac{1}{k^2} 
	+	(\tfrac{1}{16} + \tfrac{\cos 4\phi}{48})
	+	k^2(\tfrac{11}{4608} + \tfrac{\cos 4\phi}{640} + \tfrac{\cos 8\phi}{4608})
			\nonumber\\&{}
	+	k^4(\tfrac{3}{40960} + \tfrac{31 \cos 4\phi}{442368} + \tfrac{17 \cos 8\phi}{860160} + \tfrac{\cos 12\phi}{442368})
	+ \dotso
	.
	\end{align}
Take Eq.~\eqref{eqGIntegral} and extend the domain of integration to the entire $(p,q)$ plane.  Let $x=r\cos\varphi$ and $y=r\sin\varphi$.  Then
	\begin{align}
	G_{xy} &= 
	\frac{1}{(2\pi)^2}
	 \int_0^{\infty} dk~ k
	 \int_0^{2\pi} d\phi~
	 e^{i k \cos (\varphi - \phi)}
			\nonumber\\&{}
	 \bigg[
	\tfrac{1}{k^2} 
	+	(\tfrac{1}{16} + \tfrac{\cos 4\phi}{48})
	+	k^2(\tfrac{11}{4608} + \tfrac{\cos 4\phi}{640} + \tfrac{\cos 8\phi}{4608})
	+ \dotso
	\bigg]
	%\label{eqGIntegral}
	\end{align}
Now let us derive an identity for the Fourier transform of a 2D power law function,
	\begin{align}
	 %\int \frac{d^2k}{(2\pi)^2}~ 
	 &
		\frac{1}{(2\pi)^2}
	 \int_0^{\infty} dk~ k
	 \int_0^{2\pi} d\phi~
	 e^{ikr \cos (\varphi - \phi)}	 ~~
	 k^n e^{im\phi}
			\nonumber\\
	&=
		\frac{e^{im\varphi}}{4\pi^2}
	 \int_0^{\infty} dk~ k^{n+1}
	 \int_0^{2\pi} d\phi~
	 e^{ikr \cos \phi}	
			\nonumber\\
	&=
		\frac{i^m  e^{im\varphi}}{2\pi}
	 \int_0^{\infty} dk~ k^{n+1}
		J_m(kr)
			\nonumber\\
	&=
	 	\frac{2^n i^m \Gamma(1+\tfrac{m+n}{2}) }{\pi \Gamma(\frac{m-n}{2})   }
 			\frac{e^{im\varphi} }{r^{n+2}}
	.
	\label{eqFT2DPowerLaw}
	\end{align}
This gives
	\begin{align}
	2\pi G_{xy} &= 
		c_1
		+ \ln \tfrac{1}{r}
		+	\tfrac{\cos 4\phi}{12r^2}
		+	\tfrac{3\cos 4\phi}{40r^4}
		+	\tfrac{5\cos 8\phi}{48r^4}
		+ \dotso
	\end{align}
The constant $c_1$ is infinite, but the other terms are finite.  Thus the regularized Green function has the form
	\begin{align}
	2\pi F_{xy} &= 
		c_2
		+ \ln r
		-	\tfrac{\cos 4\phi}{12r^2}
		-	\tfrac{3\cos 4\phi}{40r^4}
		-	\tfrac{5\cos 8\phi}{48r^4}
		- \dotso
	\end{align}
where $c_2$ is a finite constant.
By matching this to the inverse power series for the harmonic numbers, Eq.~\eqref{eqHarmonicNumbers}, it can be shown that
	\begin{align}
	2\pi F_{xy} &= 
		\gamma + \ln \sqrt{8}	+ \ln r
		-	\tfrac{\cos 4\phi}{12r^2}
		- \tfrac{1}{r^4} \big( 
				\tfrac{3\cos 4\phi}{40}
			+	\tfrac{5\cos 8\phi}{48}
			\big)
			\nonumber\\&{}
		- \tfrac{1}{r^6} \big( 
				\tfrac{51\cos 8\phi}{112}
			+	\tfrac{35\cos 12\phi}{72}
			\big)
			\nonumber\\&{}
		- \tfrac{1}{r^8} \big( 
				\tfrac{217\cos 8\phi}{320}
			+	\tfrac{45\cos 12\phi}{8}
			+	\tfrac{1925\cos 16\phi}{384}
			\big)
			\nonumber\\&{}
		- \tfrac{1}{r^{10}} \big( 
				\tfrac{38859\cos 12\phi}{1408}
			+	\tfrac{3795\cos 16\phi}{32}
			+	\tfrac{35035\cos 20\phi}{384}
			\big)
		- \dotso
	\label{eqGreenFunctionSeries}
	\end{align}
where $\gamma\approx 0.577216$ is the Euler-Mascheroni constant.\cite{kleinertBook,atkinson1999,cserti2000}

For $r=60$, the $1/r^{10}$ terms in Eq.~\eqref{eqGreenFunctionSeries} have absolute value smaller than $10^{-16}$.  Therefore, truncating the series at the $1/r^8$ term allows us to evaluate $F_{xy}$ to machine precision for all $r>60$.  In our DLA simulations we obtain $F_{xy}$ by table lookup for $x<60$ and $y<60$ and using the series otherwise.

%============================================================================
\section{Alternative methods \label{secDifficult}}
%============================================================================
In this appendix we discuss other approaches to bias-free lattice DLA, which are less efficient than the algorithm we have presented.

%----------------------------------------------------------------------------
\myheading{Solution of Laplace's equation for an arbitrary cluster:}
%----------------------------------------------------------------------------
For a particle launched at infinity, the first-passage probabilities to the sticky sites (sites adjacent to cluster sites) can be computed exactly by solving Laplace's equation.  One can then add a particle at a position picked directly according to these probabilities.  However, for a cluster of $N$ sites, solving Laplace's equation is a dense linear algebra problem taking $O(N^3)$ time.  Even if the Sherman-Morrison formula\cite{numericalRecipes} is used to update the inverse matrix incrementally, the problem still takes $O(N^2)$ time for every particle that is added to the cluster.  This is prohibitively slow.

%----------------------------------------------------------------------------
\myheading{Walk-in-to-circle methods:}
%----------------------------------------------------------------------------
For 2D or 3D continuum DLA, it is easy to return a particle to the bounding circle or sphere of the cluster.  The electrostatic problem is easily solved by the method of images, and the return probability distribution can be evaluated and sampled analytically.  This is exploited in a killing-free algorithm for continuum DLA\cite{menshutin2006,menshutin2011} where particles that escape from the launching circle are immediately returned to the launching circle.  
For lattice DLA, however, circular or spherical boundaries do not fit naturally on the lattice.  Snapping to the grid leads to large errors, as we have shown.  It may be possible to reduce these errors by returning the particle to a fuzzy annulus; we have not investigated this completely.

	%&&&&&&&&&&&&&&&&&&&&&&&&&&&&&&&&&&&&&&&&&&&&&&&&&&&&&&&&&&&&&&&&&&&&&&&&&&&&
	% FIGURE
	%&&&&&&&&&&&&&&&&&&&&&&&&&&&&&&&&&&&&&&&&&&&&&&&&&&&&&&&&&&&&&&&&&&&&&&&&&&&&
	\begin{figure}[!htb]
		\includegraphics[width=0.6\columnwidth]{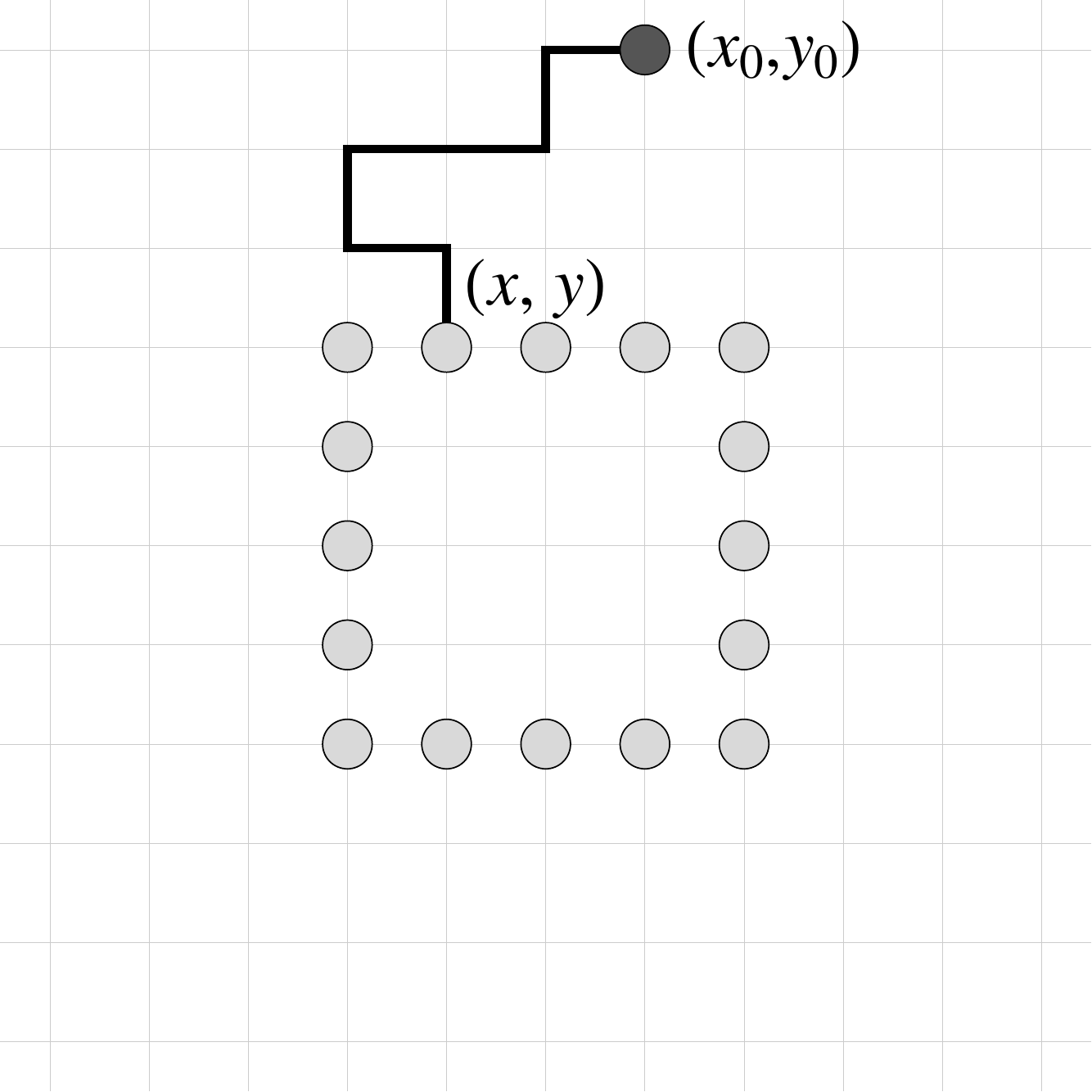}
	\caption{	
		\label{WalkInToSquare}
		Suppose a particle starts at $(x_0,y_0)$ (black disk)
			and executes a random walk on the square lattice 
			until it reaches a point $(x,y)$ on a square of side $l$ (gray disks).
		The first-passage probabilities can be found using brute-force matrix computation,
			which takes $O(l^3)$ time,
			but there is no simple analytic formula for general $l$.
	}
	\end{figure}

%----------------------------------------------------------------------------
\myheading{Walk-in-to-square methods:}
%----------------------------------------------------------------------------
What if we wish to return a particle to a bounding rectangle, square, or cube?  This requires finding the charge distribution on a conductor induced by an \emph{exterior} point charge.  
Whereas interior electrostatics problems are amenable to a variety of methods (images, separation of variables, and conformal mapping), 
exterior electrostatics problems are notoriously difficult even in the continuum.  
The most accurate results for the capacitance of a cube have been obtained by mapping the electrostatic problem back to a random walk problem and using Monte Carlo techniques!\cite{wintle2004,hwang2004,hwang2005}  It is probably futile to search for a simple analytic formula for the lattice problem (see Fig.~\ref{WalkInToSquare}).  

The walk-in-to-square problem can be solved numerically, but this takes $O(l^3)$ time, where $l$ is the perimeter of the boundary.  In comparison, the iterated walk-to-line method may take several hundred iterations to return the walker to the bounding box of the cluster, but this number of iterations is independent of cluster size.  Thus we prefer the walk-to-line method.	

%============================================================================
% Bibliography
%============================================================================

%==== If using BibTeX ====
%\bibliography{dla}

%==== If not using BibTeX ====
% I THINK PRL DOES THIS AUTOMATICALLY

\end{document}